\documentclass[lettersize,journal]{IEEEtran}

\usepackage{amsmath}
\usepackage{algorithmic}
\usepackage{algorithm}
\usepackage{array}
\usepackage[caption=false,font=small,labelfont=rm,textfont=rm]{subfig}
\usepackage{textcomp}
\usepackage{stfloats}
\usepackage{url}
\usepackage{verbatim}
\usepackage{graphicx}
	\graphicspath{{./Figures/}}
\usepackage{cuted}
\usepackage{multirow}
\usepackage{xcolor} 
\usepackage{bm,bbm}
\usepackage{threeparttable} 
\usepackage{booktabs}
\usepackage{multicol}
\usepackage{enumitem}
\usepackage{amssymb}
\usepackage{placeins} 
\usepackage[normalem]{ulem}
\usepackage{rotating}
\usepackage[numbers,compress]{natbib}
\usepackage{siunitx}
\usepackage[acronym]{glossaries}

\counterwithout{table}{section}
\makeglossaries

\newacronym{cmp}{CMP}{covariance matrix patch}
\newacronym{insar}{InSAR}{interferometric SAR}
\newacronym{rmse}{RMSE}{root-mean-square error}
\newacronym{pl}{PL}{phase-linking}
\newacronym{ad}{AD}{Anderson-Darling}
\newacronym{acaf}{ACAF}{angular consistency adaptive filter}
\newacronym{fashps}{FaSHPS}{fast statistically homogeneous pixel selection}
\newacronym{spdc}{SDPC}{similarly decorrelated pixel clustering}
\newacronym{pdf}{pdf}{probability density function}
\newacronym{ccg}{CCG}{complex circular Gaussian}
\newacronym{ds}{DS}{distributed scatterer}
\newacronym{mc}{MC}{Monte Carlo}
\newacronym{pta}{PTA}{phase triangulation algorithm}
\newacronym{emi}{EMI}{eigendecomposition-based maximum-likelihood-estimator of interferometric phase}
\newacronym{shp}{SHP}{statistically homogeneous pixel}
\newacronym{sshp}{SSHP}{shape statistically homogeneous pixel}
\newacronym{cgg}{CGG}{complex generalized Gaussian}
\newacronym{cacg}{CACG}{complex angular central Gaussian}
\newacronym{ks}{KS}{Kolmogorov–Smirnov}
\newacronym{ces}{CES}{complex elliptically symmetric}
\newacronym{mle}{MLE}{maximum likelihood estimate}
\newacronym{cggpl}{CGG-PL}{CGG-based phase-linking}
\newacronym{cfpl}{CFPL}{covariance fitting-based phase-linking}
\newacronym{ccmmtf}{CCM-MTF}{complex covariance matrix-based multitemporal filtering}
\makeatletter
\renewcommand{\maketag@@@}[1]{\hbox{\m@th\normalsize\normalfont#1}}%
\makeatother

\begin{document}
\title{Shape-to-Scale InSAR Adaptive Filtering and Phase Linking under Complex Elliptical Models}
\author{Shuyi Yao,~\IEEEmembership{Student Member,~IEEE}, Alejandro C.\ Frery,~\IEEEmembership{Fellow,~IEEE}, Timo Balz,~\IEEEmembership{Senior Member,~IEEE}
	
	\thanks{This work was supported by the National Key Research and Development Program of China under Grant 2023YFE0110400. \emph{(Corresponding author: Shuyi Yao)}.}
	\thanks{Shuyi Yao is with the State Key Laboratory of Information Engineering in Surveying, Mapping and Remote Sensing, Wuhan University, Wuhan 430079, China, and also with the School of Mathematics and Statistics, Victoria University of Wellington, New Zealand (e-mail: yaoshuyi@whu.edu.cn). 
		
		Alejandro C.\ Frery	is with the School of Mathematics and Statistics, Victoria University of Wellington, New Zealand.
		
		Timo Balz is with the State Key Laboratory of Information Engineering in Surveying, Mapping and Remote Sensing, Wuhan University, Wuhan 430079, China (e-mail: balz@whu.edu.cn).  }
}

\markboth{This manuscript has been submitted to the IEEE Transactions on Geoscience and Remote Sensing for review}%
{Shell \MakeLowercase{\textit{et al.}}: CLTC-PL: A Robust Novel Mathematical Framework and Algorithm for InSAR Phase-Linking Based on the Central Limit Theorem for Circular Data}


\maketitle

\begin{abstract}
Distributed scatterers in InSAR (DS-InSAR) processing are essential for retrieving surface deformation in areas lacking strong point targets. 
Conventional workflows typically involve selecting statistically homogeneous pixels based on amplitude similarity, followed by phase estimation under the complex circular Gaussian model. 
However, amplitude statistics primarily reflect the backscattering strength of surface targets and may not sufficiently capture differences in decorrelation behavior. 
For example, when distinct scatterers exhibit similar backscatter strength but differ in coherence, amplitude-based selection methods may fail to differentiate them. 
Moreover, CCG-based phase estimators may lack robustness and suffer performance degradation under non-Rayleigh amplitude fluctuations.

Centered around scale-invariant second-order statistics, we propose ``Shape-to-Scale,'' a novel DS-InSAR framework. 
We first identify pixels that share a common angular scattering structure (``shape statistically homogeneous pixels'') with an angular consistency adaptive filter: a parametric selection method based on the complex angular central Gaussian distribution. 
Then, we introduce a complex generalized Gaussian-based phase estimation approach that is robust to potential non-Rayleigh scattering.

Experiments on both simulated and SAR datasets show that the proposed framework improves coherence structure clustering and enhances phase estimation robustness.
This work provides a unified and physically interpretable strategy for DS-InSAR processing and offers new insights for high-resolution SAR time series analysis.
\end{abstract}


\section{Introduction}
\IEEEPARstart{T}{ime-series} interferometric synthetic aperture radar (InSAR) are powerful tools for precisely monitoring surface deformation over large areas. 
InSAR time-series analysis relies primarily on two types of scatterers for coherent phase retrieval: persistent scatterers (PS)~\cite{ferretti_nonlinear_2000,ferretti_permanent_2001}, which maintain high coherence over time and can be individually tracked, and distributed scatterers, which form statistically homogeneous clusters with partially coherent returns~\cite{ferretti_new_2011}. 
While PS-based methods are effective in urban or man-made environments, DS-based approaches extend the applicability of this technique to natural or non-urban areas where dominant reflectors are sparse.

In DS-InSAR, the phase information is often embedded in partially coherent signals, which are susceptible to speckle and temporal decorrelation. 
This necessitates statistical estimation techniques to extract reliable signals from noisy observations. 
A widely adopted model in this context is the \gls{ccg} distribution~\cite{jong-sen_lee_intensity_1994,rocca_modeling_2007,guarnieri_hybrid_2007}. 
The CCG assumption provides analytical tractability and facilitates maximum likelihood-based estimation of interferometric phases~\cite{guarnieri_exploitation_2008}.

However, SAR signals are often collected from complex and heterogeneous terrain, where the underlying scattering mechanisms violate the assumptions of the CCG model. 
In particular, the Gaussian assumption is violated in the presence of structural heterogeneity, often leading to 
heavy-tailed amplitude distributions.

To address this issue, a common strategy in DS-InSAR is the selection of statistically homogeneous pixels which are assumed to exhibit similar scattering behavior. 
By isolating them in homogeneous clusters, the adverse effects of model mismatch can be mitigated.  
Early methods based on pixel amplitude or intensity time series include the two-sample \gls{ks} test~\cite{ferretti_new_2011}, while the Anderson–Darling (AD) test was later introduced as a more appropriate choice for heavy-tailed data~\cite{wang_retrieval_2012,goel_distributed_2014}.  
The \gls{fashps} is a widely adopted and computationally efficient parametric method that exploits the intensity statistics under the \gls{ccg} assumption~\cite{jiang_fast_2015}.  
The \gls{ccmmtf}~\cite{dong_unified_2018} and \gls{cmp}~\cite{zhao_statistically_2022} methods further incorporate phase information while remaining within the parametric framework defined by the \gls{ccg} model.  
Alternatively, the two-sample Kuiper’s test~\cite{pepe_adaptive_2021} and the \gls{spdc} method~\cite{yao_phase-based_2024} offer non-parametric phase-only strategies based on directional statistics~\cite{mardia_directional_2000} for \gls{shp} selection.  
Non-local approaches assign soft labels (i.e., similarity-based weights) to neighboring pixels, thereby achieving a weighted aggregation effect that is similar to homogeneous pixel selection~\cite{deledalle_nl-insar_2011,deledalle_nl-sar_2015,aghababaei_nonlocal_2022,shen_homenl_2023}.  
More recently, deep learning-based methods have also been proposed to identify SHPs in a data-driven manner~\cite{hu_deep_2022,zhao_psmnet_2025}.

Some studies have also addressed the treatment of heterogeneous scatterers from a robust estimation perspective~\cite{vu_robust_2023}.  
By adopting robust statistical models, these methods aim to suppress the influence of amplitude fluctuations and achieve stable estimation of the second-order statistics.
Common robust models used in InSAR include the complex $t$-distribution~\cite{wang_robust_2016,zhang_robust_2024}, the complex $K$-distribution~\cite{zwieback_repeat-pass_2021}, both of which belong to the family of \gls{ces} distributions. 
These models offer analytical tractability while providing robustness against heavy-tailed amplitude fluctuations. 

Despite their flexibility and robustness, \gls{ces} models share a fundamental limitation: the entirety of their second-order behavior is still governed by a single scatter (or complex covariance) matrix $\bm{\Sigma}$, which corresponds to one specific decorrelation behavior or coherence structure. 
This unified modeling may hinder the ability to distinguish physically distinct scattering mechanisms, potentially compromising the interpretability and physical consistency of the results.

Let us reconsider the core characteristics of \gls{ds} signals in InSAR applications. The statistical behavior of interferometric phase is primarily governed by coherence, which quantifies the relative similarity between two electromagnetic signals and is independent of signal magnitude. The coherence matrix captures the mutual coherence between complex SAR signals acquired at different times and thus describes the directional preference of SAR observations in high-dimensional space, where the overall scale is fixed. As a scale-invariant second-order statistic, it plays a critical role in phase estimation. For instance, the probabilistic model of interferometric phase proposed by Lee et al.~\cite{jong-sen_lee_intensity_1994} is independent of the signal's amplitude scale.  

This scale-insensitivity motivates a shift in statistically homogeneous pixels (SHP) selection strategy: we propose to select pixels entirely based on their consistency in the directionality of scattering, which reflects a shared interferometric coherence structure and more directly targets the statistical properties most relevant to phase inference. 
As a result, the influence of potential non-Rayleigh amplitude fluctuations can be effectively suppressed, making the approach adaptable to complex land surfaces. 
Furthermore, pixels with substantial intensity differences but similar coherence structures can be grouped, enabling increased sample sizes while preserving underlying physical consistency. 
Since no intensity-based distinction is made during the SHP selection stage, the subsequent phase estimation method should either be scale-invariant or capable of flexibly adapting to varying intensity distributions.

Accordingly, we propose a novel \emph{Shape-to-Scale framework}. 
We first identify pixels that share a common angular structure, which mathematically corresponds to having the same \emph{shape matrix} and physically implies similar decorrelation behavior. 
We refer to these pixels as \glspl{sshp}. 
To achieve this, the original multi-dimensional SAR signals are projected onto the unit complex sphere, enabling statistical modeling under the \gls{cacg} distribution. 
Notably, since all \gls{ces} distributions reduce to the \gls{cacg} model after scale normalization, the \gls{cacg} model naturally summarizes all possible intensity distributions while isolating the essential structural characteristics of interferometric coherence. 
Based on this model, we develop a parametric SHP selection method, termed ``angular consistency adaptive filter'' (ACAF).
Subsequently, to account for potential scale-based non-Gaussianity, we adopt a \gls{pl} approach based on the \gls{cgg} model, which explicitly models intensity variation, thereby achieving a balance between robustness and efficiency for phase estimation.

This separation of shape and scale enables a robust, interpretable, and tractable phase estimation in \gls{ds} areas. 
Built upon the established two-step framework of \gls{shp} selection followed by interferometric phase estimation of SqueeSAR~\cite{ferretti_new_2011}, the proposed workflow can be seamlessly integrated into existing time-series InSAR processing pipelines.

This paper is organized as follows. 
Section~\ref{sec:statistical_modeling} introduces the statistical models and motivates the proposed Shape-to-Scale approach. Section~\ref{sec:method} describes the proposed SHP selection and CGG-based phase estimation methods in detail. 
Experimental results on both simulated and observed InSAR data are presented in Section~\ref{sec:sim} and~\ref{sec:real}, respectively. Section~\ref{sec:conclusion} concludes the paper.

\section{Complex Elliptical Modeling of InSAR Time Series}\label{sec:statistical_modeling}

We consider a temporal stack of $N$ co-registered single-look complex (SLC) images acquired over the same scene.
Let $\mathbf{z}_i \in \mathbbm{C}^N$ denote the temporal SAR vector of the $i$-th pixel in a local window; assume further that these $L$ observations are stationary and independent.
The set of such vectors is denoted by $\left\{ \mathbf{z}_i \right\}_{i=1}^{L}$. 
We define the corresponding observation matrix as
\[
\mathbf{Z} = 
\left[ \mathbf{z}_1,\mathbf{z}_2,\dots,\mathbf{z}_L \right] 
\in \mathbbm{C}^{N \times L},
\]
where each column represents a single pixel's time series.

A classical assumption in InSAR analysis is that $\mathbf{z}_i$ follows a zero-mean \gls{ccg} distribution:
\[
\mathbf{z}_i \sim \mathcal{CN}(\mathbf{0}, \bm{\Sigma}),
\]
where $\bm{\Sigma} \in \mathbbm{C}^{N \times N}$ is the full-rank scatter (or covariance) matrix. Under this model, each $\mathbf{z}_i$ is fully characterized by second-order statistics. 
Under the CCG model, many widely-used estimators, including the sample coherence matrix and the maximum-likelihood-based phase triangulation methods, have optimal properties.

In many InSAR time series applications, the \gls{ccg} assumption may not fully align with the characteristics of real data, especially in high-resolution SAR datasets~\cite{wang_retrieval_2012,bai_multipass_2025} or areas with heterogeneous land cover~\cite{zhang_nonlocal_2024}.
While it offers analytical tractability and has been widely adopted for its mathematical convenience, it cannot capture the heavy-tailed amplitude fluctuations and heterogeneity interferometric coherence structure found in the SAR data.

To improve model fitness and robustness in SAR and InSAR data analysis, \gls{ces} distributions can be adopted because they generalize the CCG model. 
A random vector $\mathbf{z} \in \mathbbm{C}^N$ is said to follow a zero-mean CES distribution $\mathbf{z} \sim \mathcal{CES}_g(\mathbf{0}, \bm{\Sigma})$ if its \gls{pdf} (when it exists) can be expressed as
\begin{equation}\label{eq:ces_pdf}
f_{\mathbf{z}}(\mathbf{z}) = C_{N,g} \big(\det(\bm{\Sigma})\big)^{-1}
g\left( \mathbf{z}^{\dagger} \bm{\Sigma}^{-1} \mathbf{z} \right),
\end{equation}
where $\det(\cdot)$ denotes determinant, $\bm{\Sigma} \in \mathbbm{C}^{N \times N}$ is a positive definite 
scatter matrix\footnote{Note that the scatter matrix may differ from the covariance matrix by a scale factor~\cite{ollila_complex_2012}.}, 
$g(\cdot)$ is a scalar density generator function, and $C_{N,g}$ is a normalizing constant.
Typical examples include the complex $t$-distribution, 
the complex $K$-distribution, the complex $\mathcal{G}^0$ distribution~\cite{freitas_polarimetric_2005} and the \gls{cgg} distribution, among others.
CES distributions are particularly effective in modeling heavy-tailed amplitude statistics. However, the entire second-order statistic is still governed by a single scatter matrix $\bm{\Sigma}$, which cannot distinguish between physically distinct decorrelation mechanisms. CES models tend to ``absorb'' structural heterogeneity into a single high-capacity representation, making model fitting appear statistically adequate while concealing underlying physical diversity in scattering behavior.

To address this limitation, we revisit the angular structure embedded within CES distributions. A key insight is that, when projected onto the unit complex hypersphere, i.e.,
\[
\widetilde{\mathbf{z}} = \frac{\mathbf{z}}{\|\mathbf{z}\|} \in \mathbbm{C}S^{N-1} = \left\{ \mathbf{z} \in \mathbbm{C}^N \mid \|\mathbf{z}\|_2 = 1 \right\},
\]
any zero-mean CES-distributed vector follows a \gls{cacg} distribution\, \
\[
\widetilde{\mathbf{z}} \sim \mathcal{CACG}(\bm{\Sigma}).
\]
The \gls{pdf} of the \gls{cacg} distribution is given by
\begin{equation}
	f_{\widetilde{\mathbf{z}}}(\mathbf{z}) =
	\frac{ \Gamma(N) }{ 2\pi^N \det(\bm{\Sigma}) } 
	\left( \mathbf{z}^{\dagger} \bm{\Sigma}^{-1} \mathbf{z} \right)^{-N}, \quad \text{for } \|\mathbf{z}\| = 1,
\end{equation}
where $\bm{\Sigma} \in \mathbbm{C}^{N \times N}$ is a Hermitian positive-definite matrix identifiable only up to a scale factor, as $c \bm{\Sigma}$ yields the same distribution for any $c \in \mathbbm{R}^+$.
Although the \gls{cacg} distribution has the apparent form of the density of a \gls{ces} distribution \eqref{eq:ces_pdf} by choosing
\[
g(t) = t^{-N}, \quad \text{and} \quad C_{N,g} = \frac{\Gamma(N)}{2\pi^N},
\]
the CACG distribution is \emph{not} a true member of the CES family because its support is restricted to the unit hypersphere $\mathbbm{C}S^{N-1}$. 

The CACG distribution is characterized solely by a normalized scatter matrix—referred to as the shape matrix $\mathbf{V}=N\bm{\Sigma}/\operatorname{tr}(\bm{\Sigma} )$, where $\operatorname{tr}(\cdot)$ denotes the trace—which captures the \emph{angular structure} independently of magnitude.

The concept of \emph{angular structure} arises naturally in the context of \gls{ces} distributions, where $\mathbf{z}$ can be transformed from
$\mathbf{z} = \mathcal{R} \cdot \mathbf{A} \mathbf{u}$, with $\mathcal{R} \in \mathbbm{R}^+$ a real-valued random magnitude independent of $\mathbf{Au}$, $\mathbf{A} \in \mathbbm{C}^{N \times N}$ a matrix 
such that $\mathbf{A} \mathbf{A}^{\dagger} = \bm{\Sigma}$, and $\mathbf{u} \in \mathbbm{C}S^{N-1}$ a unit-norm vector uniformly distributed on the complex unit hypersphere.
The vector 
$\widetilde{\mathbf z}=\mathbf A\mathbf u / \|\mathbf A\mathbf u\|$ 
follows a CACG distribution, whose support is the complex unit hypersphere 
$\mathbbm{C}S^{N-1}$.  Its non‐uniform angular density encodes the ellipsoidal geometry of $\bm\Sigma$.

The term \emph{angular structure} emphasizes that it models the angular preference or distribution independently of the magnitude. Only after being radially scaled by the random magnitude $\mathcal{R}$ does it extend to the full complex space $\mathbbm{C}^N$, resulting in the complex elliptically symmetric vector $\mathbf{z}$. Since the angular structure of $\widetilde{\mathbf{z}}$ is independent of the overall scale of $\bm{\Sigma}$, it is fully characterized by the shape matrix $\mathbf{V}$ or, equivalently, by the scale-invariant scatter matrix $\bm{\Sigma}$.

\begin{figure}[h]
	\centering
	\includegraphics[width=3.5in]{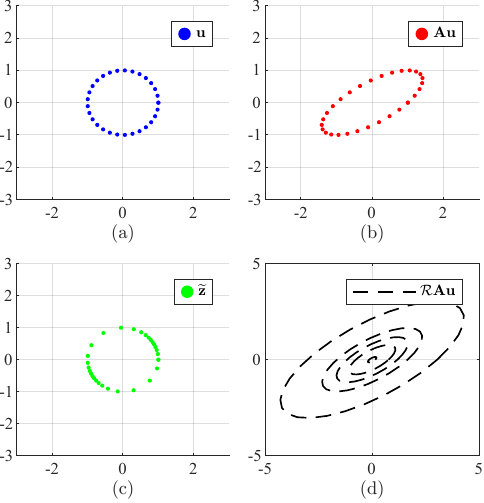} 
	\caption{Illustration of the \gls{ces} and \gls{cacg} models through a two-dimensional real-valued example. 
		(a) Unit-norm vectors $\mathbf{u} \in \mathbbm{R}S^1$ uniformly distributed on the unit circle, representing isotropic angular structure. 
		(b) Transformed vectors $\mathbf{A}\mathbf{u}$, with $\mathbf{A}\mathbf{A}^\dagger = \bm{\Sigma}$, lie on an ellipse and exhibit anisotropic angular structure determined by the scatter matrix. 
		(c) The normalized vectors $\widetilde{\mathbf{z}} = \mathbf{A} \mathbf{u} / \|\mathbf{A} \mathbf{u}\|$ follow a \gls{cacg} distribution, remaining on the unit circle but with a non-uniform angular density. 
		(d) Full CES samples $\mathbf{z} = \mathcal{R} \cdot \mathbf{A} \mathbf{u}$ are generated by radial scaling with an independent random magnitude $\mathcal{R}$ (e.g., Rayleigh distributed), filling the entire real plane.}
	\label{fig:Visual_ell} 
\end{figure}

We visualize the concept of angular structure through a simple two-dimensional real-valued example (see Fig.~\ref{fig:Visual_ell}). 
Let $\mathbf u\in S^{1}\subset\mathbb R^{2}$ be uniformly distributed on the unit circle (panel~(a) shows the quantiles uniformly spaced). 
Applying a linear transformation $\mathbf{A}\mathbf{u}$ maps the uniform directions on the unit circle into points on an ellipse (panel~(b)).
Normalizing these transformed points back to the unit circle results in a non-uniform point density on the unit hypersphere, i.e., a non-uniform angular density (panel~(c)). This non-uniformity represents the angular structure, which is determined by the scale-invariant transformation $\mathbf{A}=\bm{\Sigma}^{1/2}$. 
Finally, multiplying by an independent random radius $\mathcal{R}$ generates the full CES samples (panel~(d)), filling the entire two-dimensional real space according to the distribution of $\mathcal{R}$, such as the Rayleigh distribution in the case of \gls{ccg} distribution.

In summary, the angular structure characterizes the preferred directions of multidimensional SAR data on the complex unit hypersphere and is therefore inherently linked to the intrinsic \emph{decorrelation behavior} among interferograms in \gls{ds} areas.  
Specifically, projecting any two dimensions of the $N$-dimensional vector $\widetilde{\mathbf z}$ onto the complex plane yields a two-dimensional slice whose angular density encodes the scale-invariant complex covariance between the corresponding SAR acquisitions.  
Hence, the angular structure can be viewed as the ensemble of all such pairwise, scale-invariant complex covariances, which is fully determined by the normalized scatter (shape) matrix $\mathbf{V}$.  
Note that $\mathbf{V}$ is not identical to the coherence matrix $\bm \Gamma$: it only fixes the overall scale.  
However, under the assumption that each SAR acquisition has equal signal energy, $\mathbf{V}$ coincides with the coherence matrix.

This property allows us to separate the tasks of identifying decorrelation behaviors through \ SHP selection and addressing potentially heavy-tailed magnitude fluctuations in the phase estimation stage. 
By first selecting pixels that share the same angular structure, we not only preserve a physically consistent decorrelation pattern, but also effectively mitigate the tendency of CES models to absorb heterogeneous pixels due to their high modeling flexibility.
In addition, a small portion of erroneously selected SSHPs can still be absorbed by CES-model-based phases estimation methods, preserving both fitting capability and estimation robustness.

\section{Preposed Shape-to-Scale apporath}\label{sec:method}

\subsection{SSHP selection based on the angular structure}

\subsubsection{The test statistic}
We developed a parametric CACG-based \gls{acaf} to identify \glspl{sshp}. Given an initial set of unit-norm temporal vectors $\left\{ \widetilde{\mathbf{z}}_i \right\}_{i=1}^L$, we aim to identify a statistically homogeneous subset whose angular structure is governed by a common scale-invariant scatter matrix $\bm{\Sigma}$. This scatter matrix $\bm{\Sigma}$ is estimated by maximizing the likelihood under the CACG model~\cite{gini_covariance_2002}, using the following fixed-point iteration (Tyler's $M$-estimator~\cite{tyler_distribution-free_1987}), for $k=0,1,\dots$:
\begin{equation}\label{eq:tyler}
	\begin{aligned}
		\widetilde{\bm{\Sigma}}^{(k+1)} &= \frac{N}{L} \sum_{i=1}^{L} 
		\frac{{\mathbf{z}}_i {\mathbf{z}}_i^{\dagger}}
		{ {\mathbf{z}}_i^{\dagger} ( \widehat{\bm{\Sigma}}^{(k)} )^{-1} {\mathbf{z}}_i }, \\[6pt]
		\widehat{\bm{\Sigma}}^{(k+1)} &= \frac{N}{\operatorname{tr}( \widetilde{\bm{\Sigma}}^{(k+1)} )} \, \widetilde{\bm{\Sigma}}^{(k+1)} ,
	\end{aligned}
\end{equation}
where we used $\widehat{\bm{\Sigma}}^{(0)}$ is the identity matrix.
Since the estimator is scale-invariant, the vectors $\mathbf{z}_i$ need not be explicitly normalized to $\widetilde{\mathbf{z}}_i$ for the computation of $\widehat{\bm{\Sigma}}$.

Then, for each candidate sample $\widetilde{\mathbf{z}}$, we compute the CACG-based test statistic $t$:
\begin{equation}\label{eq:t}
	t(\widetilde{\mathbf{z}}; \widehat{\bm{\Sigma}}) = \widetilde{\mathbf{z}}^{\dagger} \widehat{\bm{\Sigma}}^{-1} \widetilde{\mathbf{z}}. 
\end{equation}
Assume that the reference angular structure has been established, with its scatter matrix denoted by $\bm{\Sigma}_{\text{ref}}$ (scale-invariant). Let the modulus-normalized vector corresponding to the pixel under test be denoted as $\widetilde{\mathbf{z}}$.
We are interested in verifying the null hypothesis
$$ \mathbbm{E}\big[ \widetilde{\mathbf{z}} \widetilde{\mathbf{z}}^\dagger \big] = \frac{N}{\operatorname{tr}(\bm{\Sigma}_{\text{ref}})} \bm{\Sigma}_{\text{ref}} $$
because it establishes that the shape matrices are equal.
Under the null hypothesis that $\widetilde{\mathbf{z}}$ follows a CACG distribution with the same angular structure as the reference set, i.e., governed by the same scale-invariant scatter matrix $\bm{\Sigma}_{\text{ref}}$, the test statistic $t$ follows a certain distribution. 
The empirical distribution of $t$ is efficiently estimated via \gls{mc} parametric bootstrap by generating $\widetilde{\mathbf{z}} = \mathbf{A_{\text{ref}}} \mathbf{u} / \| \mathbf{A_{\text{ref}}} \mathbf{u} \|$, where $\mathbf{A}_{\text{ref}} = \bm{\Sigma}_{\text{ref}}^{1/2}$.

Therefore, we exclude angularly inconsistent samples based on the $t$-statistic and re-estimate the scatter matrix $\widehat{\bm{\Sigma}}$ using the remaining samples. This procedure is iterated until convergence.
 
\subsubsection{Initial \gls{sshp} set derived from temporal autocorrelation functions}
To ensure stable convergence, we do not use all $L$ samples for the initial estimation of the scatter matrix. 
Instead, we construct an initial \gls{sshp} set by selecting a subset of pixels based on their temporal autocorrelation functions.

Since $\widetilde{\mathbf{z}}$ (and $\mathbf{z}$) exhibit random phase offsets across acquisitions (e.g., due to atmospheric delays), 
we first align their interferometric phases to a common near-zero-mean baseline before computing the temporal autocorrelation functions.
We extract a reference global phase alignment from the scatter matrix $\widehat{\bm{\Sigma}}$, 
estimated from all $L$ samples using \eqref{eq:tyler}, as follows:

\begin{enumerate}
	\item Perform eigen-decomposition on $\widehat{\bm{\Sigma}}$ and extract the eigenvector $\mathbf{v}_{\text{max}}$ associated with the largest real eigenvalue.
	\item Construct a phase correction vector $\exp(-\mathrm{j} \angle(\mathbf{v}_{\text{max}}))$ and apply it to all $\widetilde{\mathbf{z}}_i$ via element-wise multiplication.
\end{enumerate}

The resulting aligned vector sample has its global phase fluctuation roughly removed, enabling a meaningful temporal autocorrelation estimate.

We compute the autocorrelation sequence over a set of $N_{\mathrm{lags}}$ non-negative lags $\{\tau_1, \dots, \tau_{N_{\mathrm{lags}}}\}$. Specifically, we define the empirical autocorrelation at lag $\tau$ as
\begin{equation}\label{autocorr}
R_\tau(i) = \left|\sum_{n=1}^{N - \tau} \widetilde{z}_i(n) \, \overline{\widetilde{z}_i(n + \tau)}\right|,\quad i = 1, \dots, N_{\mathrm{lags}},
\end{equation}
where $\widetilde{z}_i(n)$ denotes the $n$-th entry of $\widetilde{\mathbf{z}}_i$. For each sample $i$, the autocorrelation values satisfy that $R_\tau(i) \in [0, 1]$, with $R_0(i) = 1$, as each $\widetilde{\mathbf{z}}_i$ has unit energy. 
\subsubsection{The complete \gls{sshp} selection procedure}
Although the standard strategy would be to start selecting directly around the reference pixel at the center of the window, we use a different approach.
A selection strategy around the reference pixel may begin by identifying a subset of pixels whose temporal autocorrelation sequences closely match that of the reference pixel (e.g., based on Euclidean distance).
This subset can then be refined by iteratively applying a two-sided statistical test on the $t$-statistic to exclude dissimilar samples, continuing the process until convergence.
However, this strategy has two key drawbacks.
First, it relies on comparisons between the autocorrelation sequences of different pixels, which can amplify the estimation errors in these functions. In particular, the strategy is highly sensitive to errors in the autocorrelation estimate of the reference (center) pixel~\cite{yu_wsht_2025}.
Second, the use of a two-sided statistical test results in a loss of statistical power.
As will be shown in Section~\ref{sec:sim_t}, the left-side threshold of the test statistic $t$ is less sensitive to heterogeneous pixels, limiting the test's ability to reject dissimilar samples.

We adopt a layered selection strategy. 
First, we select the top $N+1$ pixels with the highest mean temporal autocorrelation values $\overline{R_\tau}$, forming an initial set of candidates that are most likely to exhibit high phase quality.
With them, we estimate the scatter matrix $\bm{\Sigma}$ with~\eqref{eq:tyler}. 
Based on this estimate, we perform a right-sided test on the $t$-statistic as follows:
\[
\widetilde{\mathbf{z}} \text{ is retained if }  
t(\widetilde{\mathbf{z}}; \widehat{\bm{\Sigma}}) \leq q_{1-\alpha}.
\]
where $\alpha$ is the significance level, and $q_{1-\alpha}$ denotes the $(1 - \alpha)$ quantile obtained via parametric bootstrap sampling of $t$.
After the iteration converges, we obtain the most coherent group of pixels with a consistent angular structure.

We then check whether the reference pixel is included in the current mask. 
The reference pixel is considered part of the mask, if the following empirically defined condition is satisfied:
\begin{align}\label{eq:inverse}
	&\left(
	\left[
	\text{the reference pixel is selected}
	\;\land\;
	\left( \sum_{i \in \mathcal{N}} m_i \geq 3 \right)
	\right]
	\right. \nonumber \\
	&\quad\left.
	\lor\;
	\left[
	\sum_{i \in \mathcal{N}} m_i \geq 5
	\right]
	\right)
	\;\land\;
	\left(
	\frac{n_4}{\sum m_i} > 0.2
	\right),
\end{align}
where $\mathcal{N}$ denotes the $3 \times 3$ neighborhood centered at the reference pixel (including the center itself), 
$m_i \in \{0,1\}$ indicates whether pixel $i$ is selected, and $n_4$ is the number of pixels that are \num{4}-connected to the reference pixel within the selected mask. 
The purpose of \eqref{eq:inverse} is to assess whether the reference pixel is meaningfully integrated into the current mask, rather than being isolated or weakly connected.

If the reference pixel is considered part of the mask, the selection process terminates. 
Otherwise, the previously selected pixels are excluded, and a new selection is performed; we call this ``mask reversal.'' 
This process repeats until the mask includes the reference pixel. If, at any point, the number of remaining pixels is too small (e.g., fewer than $N+1$), or a certain phase quality metric is too low (e.g., the mean coherence falls below $0.15$), all remaining candidates are selected as the final SSHP group to preserve minimal statistical support.

After the iterative selection stops, if no mask reversal has been triggered, the \num{4}-connected region containing the reference pixel is extracted, followed by a refinement step.
A two-sided CACG test is repeated once. 
If the \num{4}-connected component surrounding the reference pixel is deemed too small, a smaller auxiliary window centered on the reference pixel and sized relative to the original window and data dimension $N$ (e.g., $5 \times 5$) is merged into the mask prior to testing.

This refinement step aims to remove potentially heterogeneous pixels that may remain under the following conditions:  
(1)~the mask inversion criterion was not triggered;  
(2)~the reference pixel exhibits low coherence; and  
(3)~only a small number of high-coherence pixels are present, typically near the window boundaries.  

Thus, our overall \gls{acaf} procedure is implemented through the main module in Algorithm~\ref{alg:refine_sshp}, which iteratively refines the pixel mask based on the CACG-based test statistic $t$.
This module is invoked through the following workflow to form the complete SSHP selection procedure:

\begin{enumerate}
	\item Obtain the phase-aligned unit-norm vectors $\{\widetilde{\mathbf{z}}_i\}_{i=1}^L$ and compute the sample mean autocorrelation values $\overline{R_\tau}$ for each pixel.
	
	\item Select the top $N+1$ pixels with the highest $\overline{R_\tau}$ to form an initial set $\mathcal{S}^{[0]}$ and estimate the initial scatter matrix $\widehat{\bm{\Sigma}}^{[0]}$ with \eqref{eq:tyler}. 
	Note that square brackets are used to denote the iteration index of refining 
	the SSHP set in Algorithm~\ref{alg:refine_sshp}, 
	rather than the iterations of the $M$-estimation procedure in~\eqref{eq:tyler}.
	
	\item Perform iterative angular refinement using Algorithm~\ref{alg:refine_sshp} with a one-sided (right-sided) test.
	
	\item Conduct a mask reversal check based on equation~\eqref{eq:inverse}. If the reversal condition is met, set $\overline{R_\tau} = 0$ for all pixels currently selected, and return to Step 2.
	
	\item If no mask reversal has been triggered, extract the \num{4}-connected region around the reference pixel as the initial mask, estimate $\widehat{\bm{\Sigma}}^{[0]}$, and run Algorithm~\ref{alg:refine_sshp} with a two-sided test and $K_{\max} = 1$.

	\item Output the final SSHP mask $\mathcal{S}_{\text{final}}$.
\end{enumerate}

\begin{algorithm}[ht]
	\caption{Iterative Refinement of \gls{sshp} via CACG-based statistic $t$}
	\label{alg:refine_sshp}
	\begin{algorithmic}[1]
		\REQUIRE 
		Initial unit-norm samples $\{\widetilde{\mathbf{z}}_i\}_{i=1}^L$; \\
		Significance level $\alpha$ (default $\alpha=0.05$); \\
		Maximum iteration count $K_{\mathrm{max}}$ (default $K_{\mathrm{max}}=10$); \\
		Initial shape matrix $\widehat{\bm{\Sigma}}^{[0]}$; \\
		Test type: \texttt{single} or \texttt{double}
		
		\ENSURE Refined \gls{sshp} index set $\mathcal{S}_{\text{SSHP}}$
		\vspace{1mm}		
		\FOR{$k = 1$ to $K_{\mathrm{max}}$}
		\STATE Compute test statistic $t_i = \widetilde{\mathbf{z}}_i^{\dagger} (\widehat{\bm{\Sigma}}^{[k-1]})^{-1} \widetilde{\mathbf{z}}_i$ for all $i$
		
		\STATE Estimate threshold(s) $q$ via Monte Carlo parametric bootstrap over $\mathcal{S}^{[k-1]}$
		
		\IF{test type = \texttt{single}}
		\STATE $\mathcal{S}^{[k]} = \big\{ i \in \mathcal{S}^{[k-1]} \mid t_i \leq q_{1-\alpha}^{[k]} \big\}$
		\ELSIF{test type = \texttt{double}}
		\STATE $\mathcal{S}^{[k]} = \big\{ i \in \mathcal{S}^{[k-1]} \mid q_{\alpha/2}^{[k]} \leq t_i \leq q_{1-\alpha/2}^{[k]} \big\}$
		\ENDIF
		
		\STATE Remove indices with zero autocorrelation:
		\[
		\mathcal{S}^{[k]} \leftarrow \big\{ i \in \mathcal{S}^{[k]} \mid \overline{R_\tau} > 0 \big\}
		\]
		
		\STATE Re-estimate shape matrix $\widehat{\bm{\Sigma}}^{[k]}$ from $\{\widetilde{\mathbf{z}}_i\}_{i \in \mathcal{S}^{[k]}}$ using~\eqref{eq:tyler}
		
		\STATE Compute relative change:
		\[
		\delta^{[k]} = \frac{\| \widehat{\bm{\Sigma}}^{[k]} - \widehat{\bm{\Sigma}}^{[k-1]} \|_F}{\| \widehat{\bm{\Sigma}}^{[k-1]} \|_F}
		\]
		\IF{$\delta^{[k]} < \epsilon$}
		\STATE \textbf{Break} \hfill \COMMENT{Convergence reached}
		\ENDIF
		\ENDFOR
		\STATE \textbf{Return} $\mathcal{S}_{\text{SSHP}} = \mathcal{S}^{[k]}$
	\end{algorithmic}
\end{algorithm}

For a detailed mathematical analysis of the behavior of the $t$-statistic and a formal justification for the reliability of using a right-tail test to reject low-quality, heterogeneous pixels, please refer to Appendix~\ref{appendix:t_statistic_analysis}.

\subsection{Phase history estimation via CGG model}

Based on angular structure similarity, the identified SSHPs correspond to pixels that share a physically consistent decorrelation behavior.
However, due to intrinsic scattering properties or limitations in the selection process, their amplitudes may depart from the Rayleigh distribution.
Such lack of fit highlights the need for phase estimation methods that are either scale-invariant, robust to amplitude outliers, or capable of capturing scale-based non-Gaussianity.
Relevant approaches include phase-only \gls{pl} methods~\cite{yao_phase-based_2024,pepe_improved_2015,yao_new_2024,vu_covariance_2025}, 
the covariance matrix fitting \gls{pl} method based on Tyler's estimator~\cite{vu_covariance_2025}, 
and robust modeling using complex $K$-~\cite{zwieback_repeat-pass_2021}, complex $t$-distributions~\cite{wang_robust_2016,zhang_robust_2024} or other compound models~\cite{vu_robust_2023}.

To explicitly account for the scale-based non-Gaussianity that may persist within the selected SSHPs, we introduce an alternative phase estimation method based on the \gls{cgg} model.
The \gls{cgg} distribution is a member of the \gls{ces} family that uses a single shape parameter $s > 0$ to control the non-Rayleigh amplitude fluctuations among \glspl{sshp}.
Compared with other CES models—such as the complex $t$- and complex $K$-distributions---and with non-CES alternatives like the complex $\alpha$-stable family, the CGG distribution offers superior analytical tractability.
In particular, its closed-form log-likelihood expression facilitates inference and parameter estimation under maximum likelihood principles.
While marginal amplitude statistics under CGG remain analytically intractable---which might limit its use in conventional amplitude-based SAR applications---the model is convenient for multi-dimensional phase statistics modeling in 
time-series InSAR analysis.

The \gls{pdf} of the \gls{cgg} distribution is obtained with the density generator $g(\cdot)$ (see~\eqref{eq:ces_pdf}) defined as
\[
g(t) = \exp\left( -\frac{t^s}{b} \right),
\]
where $s > 0$ is a shape parameter, and $b$ is given by:
\[
b = \left( \frac{ N \, \Gamma\left( \frac{N}{s} \right) }{ \Gamma\left( \frac{N+1}{s} \right) } \right)^s .
\]
The resulting normalizing constant in~\eqref{eq:ces_pdf} is
\[
C_{N, g} = \frac{s \, \Gamma(N) \, b^{-N/s}}{\pi^N \, \Gamma(N/s)}.
\]
Thus, the complete expression of the \gls{cgg} pdf is
\begin{equation}\label{eq:cgg_pdf}
	f_{\mathbf{z}}(\mathbf{z}) =
	\frac{ s \, \Gamma(N) \, b^{-N/s} }{ \pi^N \, \Gamma\left( \frac{N}{s} \right) \det(\bm{\Sigma})} 
	\exp\left(
	- \frac{ \left( \mathbf{z}^{\dagger} \mathbf{\Sigma}^{-1} \mathbf{z} \right)^s }{ b }
	\right) .
\end{equation}

\begin{figure}[hbtp]
	\centering
	\includegraphics[width=3.3in]{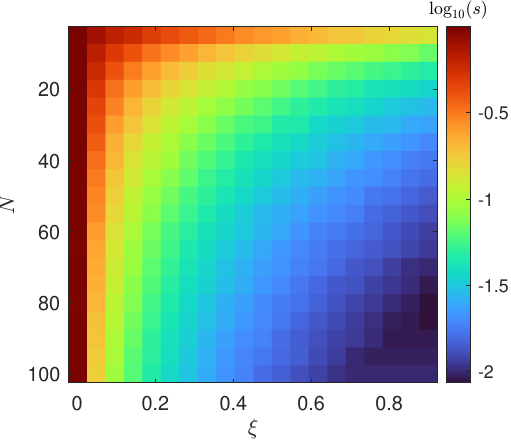} 
	\caption{CGG shape parameter $s$ (log-scale) as a function of the number of acquisitions $N$ and the variance $\xi$ of the real-valued Gamma-distributed texture.}
	\label{fig:s_behavior} 
\end{figure}

While the parameter $s$ controls the tail behavior of the CGG distribution, its interpretation is not straightforward due to the lack of a closed-form expression for the kurtosis, which is also influenced by the dimension $N$. 
To gain an intuitive understanding of how $s$ reflects in tail heaviness, we conduct a simulation study based on a product SAR model, where a \gls{ccg} vector is modulated by a Gamma-distributed texture with unit mean and variance $\xi$. 
This yields a complex $K$-distributed signal whose tail behavior is governed by $\xi$. 
For each pair of $(N,\xi)$ values, we generate $10^4$ samples and estimate the CGG shape parameter $s$. 
The resulting heat map (Fig.~\ref{fig:s_behavior}) shows that, in average, the estimated $s$ decreases monotonically with increasing $\xi$, and for a fixed $\xi$, larger $N$ also leads to smaller $s$. 
This empirical study provides a practical reference for interpreting $s$ in real InSAR applications.

We first estimate the scatter matrix $\bm{\Sigma}$ and the shape parameter $s$, which are then used as plug-in estimates for subsequent phase estimation.
\subsubsection{Estimation of $\bm{\Sigma}$ and $s$ under the CGG model}
To jointly estimate the shape parameter $s$ and the scatter matrix $\bm{\Sigma}$ under the CGG model, 
we adopt an alternating maximum likelihood estimation strategy~\cite{frery_analysis_2004}. 

We first compute an initial estimate of the scatter matrix $\widehat{\bm{\Sigma}}$ 
using the sample covariance with the selected \gls{sshp} pixels. 
Then, given the current estimate of $\bm{\Sigma}$, we update the shape parameter $s$ by maximizing the log-likelihood:
\begin{multline}\label{eq:cggd_shape_mle}
	\widehat{s} = \arg\max_{s \in \mathbbm{R}^+} \left\{
	\log s - \log \Gamma\left( \frac{N}{s} \right) 
	- \frac{N}{s} \log b \right. \\[4pt]
	\left. \quad
	- \frac{1}{L} \sum_{i=1}^{L} 
	\frac{ \left( \tilde{\mathbf{z}}_i^{\dagger} \widehat{\bm{\Sigma}}^{-1} \tilde{\mathbf{z}}_i \right)^s }
	{ b } \vphantom{ \log \Gamma\left( \frac{N}{s} \right) }
	\right\} .
\end{multline}
Subsequently, fixing the estimated $s$, we update the scatter matrix estimate using the MLE iteration under the CGG model:
\begin{equation}
	\widehat{\bm{\Sigma}}^{(k+1)} = 
	\frac{1}{L} \sum_{i=1}^L 
	\varphi\left( \mathbf{z}_i^{\dagger} ( \widehat{\bm{\Sigma}}^{(k)} )^{-1} \mathbf{z}_i \right) 
	\mathbf{z}_i \mathbf{z}_i^{\dagger},
\end{equation}
where
\[
\varphi(t) = -\frac{g'(t)}{g(t)}=\frac{s}{b} t^{s - 1}.
\]
	
These two steps are applied iteratively until the successive variations in both $s$ and $\bm{\Sigma}$ become sufficiently small, yielding a joint estimate of $(\widehat{s}, \widehat{\bm{\Sigma}})$.
Once the parameters $(\widehat{s}, \widehat{\bm{\Sigma}})$ are determined, the phase history $\widehat{\bm{\theta}}$ can be estimated using two approaches: 
(a)~the plug-in maximum likelihood estimator~\cite{ferretti_new_2011,guarnieri_exploitation_2008} under the CGG model (CGG-MLE), 
and (b)~the plug-in complex \gls{cfpl}~\cite{vu_covariance_2025,bai_lamie_2023} (CGG-CFPL). To balance the energy differences across acquisitions, the estimated scatter matrix is normalized to obtain the complex coherence matrix $\bm{\Gamma}=(\Gamma_{ij})$:
\begin{equation}\label{eq:cohm}
	\Gamma_{ij} = \frac{\Sigma_{ij}}{\sqrt{\Sigma_{ii} \Sigma_{jj}}}.
\end{equation}
Coherence provides a more reliable measure of relative differences between acquisitions, as it is independent of the absolute echo magnitude~\cite{woodhouse_introduction_2017}.

\subsubsection{Plug-in maximum likelihood estimation of phase history $\widehat{\bm{\theta}}$ (CGG-MLE)}
In CGG-MLE,
the phase history $\widehat{\bm{\theta}}$ is estimated
by maximizing the log-likelihood with respect to $\bm{\theta}$:
\begin{equation}\label{eq:cggd_phase_objective}
\widehat{\bm{\theta}}_{2:N} = \arg\max_{\bm{\theta}_{2:N}  \in \mathbbm{R}^{N-1}} \left\{ -\sum_{i=1}^{L}  
	\left\| 
	\widehat{\mathbf{G}}^{-1/2} \bm{\Theta}^{\dagger} \mathbf{z}_i 
	\right\|_2^{2{\widehat{s}}}
	\right\} ,
\end{equation}
where $\bm{\Theta}=\operatorname{diag}\{1,\exp(\mathrm{j}\theta_2), \exp(\mathrm{j}\theta_3), \dots, \exp(\mathrm{j}\theta_{N}) \}$ is a $N \times N$ diagonal matrix and $\widehat{\mathrm{G}}_{ij}=|\widehat{\Gamma}_{ij}|$. 
We denote by $\bm{\theta}_{2:n} = [\theta_2,\dots,\theta_n]^\top$ the vector consisting of the 2nd to the $n$th elements of $\bm{\theta}$.
The optima can be found using the Broyden–Fletcher–Goldfarb–Shanno (BFGS) algorithm~\cite{ferretti_new_2011}.
The complete $N$-dimensional estimated phase history is then obtained by inserting the zero-reference phase $\widehat{\bm{\theta}} = [0, \widehat{\bm{\theta}}_{2:N}^\top]^\top$.

\subsubsection{Plug-in complex covariance fitting method for phase linking (CGG-CFPL)}
In CGG-CFPL, the phase history $\widehat{\bm{\theta}}$ is estimated by minimizing the Frobenius norm between the estimated complex coherence matrix $\widehat{\bm{\Gamma}}$ and its parameterized representation based on $\bm{\theta}$:
\begin{equation}\label{eq:cggd_phase_objective_mm}
	\widehat{\bm{\theta}} = \arg\min_{\bm{\theta} \in \mathbbm{R}^{N}} \left\| \widehat{\mathbf{G}} \circ \left(\mathbf{w}_{\bm{\theta}}\mathbf{w}^\dagger_{\bm{\theta}}\right) - \widehat{\bm{\Gamma}} \right\|_F,
\end{equation}
where $\circ$ denotes the Hadamard product and  $\mathbf{w}_{\bm{\theta}}=[\exp(\mathrm{j}\theta_1), \exp(\mathrm{j}\theta_2), \dots, \exp(\mathrm{j}\theta_{N})]^\top$. This optima can be efficiently found via a majorization–minimization (MM) algorithm~\cite{vu_covariance_2025}. To ensure a zero-reference at the first acquisition, the estimated phase history is realigned as $\widehat{\bm{\theta}} \leftarrow \arg[e^{\mathrm{j}(\widehat{\bm{\theta}}-\widehat{\theta_1})}]$.

CGG-MLE is based on maximum likelihood estimation and is theoretically advantageous due to its statistical grounding, provided that the inverse of coherence magnitude matrix is accurately estimated. 
In contrast, CFPL avoids matrix inversion, which can be particularly beneficial, especially when processing high-dimensional datasets~\cite{bai_lamie_2023}.

\section{Simulation Results}\label{sec:sim}

\subsection{Statistical power of the CACG-based test statistic $t$}\label{sec:sim_t}

We evaluate the statistical power (Type~II error) of the CACG-based test statistic $t$ under variations in the coherence magnitude. 
The significance level is fixed at $\alpha = 0.05$ for all experiments.

To simulate temporal decorrelation, we adopt an exponential model of the form~\cite{rocca_modeling_2007,morishita_temporal_2015}
\begin{equation}
	\gamma(\Delta T) = p_{\rm const} + (1 - p_{\rm const}) \cdot \exp\left( -\frac{\Delta T}{2\tau} \right),
\end{equation}
where $\tau$ is the exponential decay constant, representing the time required for the coherence to decay to approximately \SI{60}{\percent} of its initial value. 
The parameter $p_{\rm const} \in [0,1]$ denotes the proportion of fully coherent elementary scatterers, while the remaining $(1 - p_{\rm const})$ fraction is assumed to follow the exponential decay.

\begin{figure}[hbtp]
	\centering
	\includegraphics[width=3.5in]{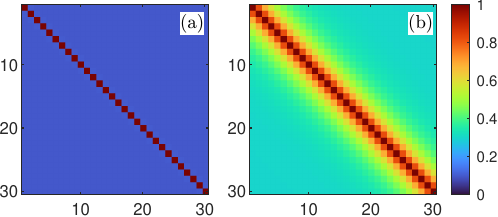} 
	\caption{
		(a)~Coherence matrix for the lowest coherence level in the experimental design.  
		(b)~Coherence matrix for the highest coherence level, representing a medium coherence scenario in the design.  
	}
	\label{fig:cohms_eachEnd} 
\end{figure}

By varying $p_{\rm const}$ from $0.1$ to $0.3$ and $\tau$ from $1$ to $20$, we generate a series of coherence magnitude matrices with increasing average coherence, ranging from low to medium levels. The scenarios corresponding to the lowest and highest coherence in the experimental design are shown as two representative cases in Fig.~\ref{fig:cohms_eachEnd}.

\begin{figure}[hbtp]
	\centering
	\includegraphics[width=3.5in]{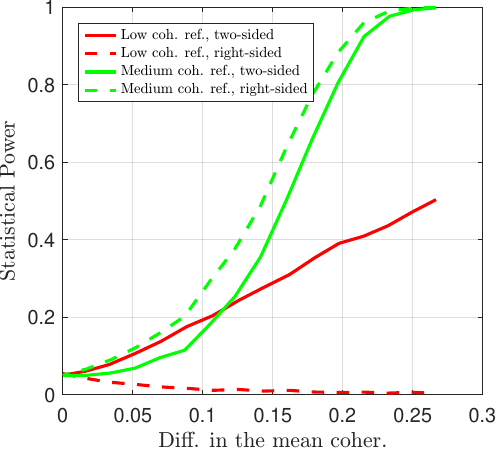} 
	\caption{Statistical power of the CACG-based $t$-test versus the difference in mean coherence induced by varying $(\tau, p_{\rm const})$.  
		Results are shown for cases where the reference pixel has low coherence (red) or medium coherence (green); solid lines indicate two-sided tests and dashed lines indicate right-sided tests. The significance level is fixed at 0.05.
	}
	\label{fig:power_mean_coh} 
\end{figure}

As shown in Fig.~\ref{fig:power_mean_coh}, when the reference pixel has an overall low coherence level (quantified by the mean coherence), the right-sided test fails to distinguish heterogeneous pixels with higher coherence. Even with the two-sided test, the statistical power remains limited: at the largest coherence difference designed in this experiment, only about half of the heterogeneous pixels are correctly rejected.
In contrast, when the reference pixel has a higher coherence level than the heterogeneous pixels, the test power increases rapidly with the coherence difference. Using a right-sided test further improves the detection power under the same false alarm rate (significance level).
This validates the proposed SSHP selection strategy: first extract the most coherent group of pixels within the window, then perform tests specifically targeting lower-coherence heterogeneous pixels, and rely on the mask inversion mechanism to ensure that the reference pixel is assigned to the correct group.

\subsection{Simulated 2-D scene test}

\begin{figure}[hbtp]
	\centering
	\includegraphics[width=3.5in]{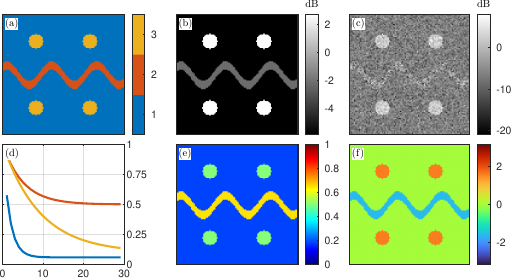} 
	\caption{
		Spatial distribution of simulation parameters:  
		(a) Classification label map based on the assumed scatterer classes;  
		(b) Class-specific mean backscatter power $\sigma^2$;  
		(c) Temporal mean intensity;  
		(d) Coherence decay curves as a function of temporal baseline for the different assumed scatterer classes (blue, red, and yellow correspond to classes 1, 2, and 3 in (a), respectively);
		(e) True mean coherence;  
		(f) True interferometric phase for the longest temporal baseline.
	}
	\label{fig:sim_settings} 
\end{figure}

We simulated a SAR time series consisting of $N = 30$ images to test the performance of proposed methods. 
The configuration of the simulated 2-D scene is illustrated in Fig.~\ref{fig:sim_settings}.
The scene contains three distinct classes of scatterers, spatially distributed as shown in Fig.~\ref{fig:sim_settings}(a). 
The complex SAR signal $\mathbf{z}$ for each pixel is generated according to the stochastic representation
\[
\mathbf{z} = \mathcal{R} \, \mathbf{A} \mathbf{u},
\]
recall that $\bm{\Sigma} = \mathbf{A} \mathbf{A}^\dagger$, $\mathcal{R}$ models the random magnitude fluctuations, and $\mathbf{u}$ is a random vector uniformly distributed on the complex unit hypersphere.
 
To simulate realistic surface complexity and non-Gaussian behavior, the amplitude $\mathcal{R}$ is assumed to follow a compound distribution: its square $\mathcal{R}^2$ follows a $K$-distribution, modeled as an exponentially distributed base modulated by an independent Gamma random variable. The expectation of $\mathcal{R}$ is fixed to $1$ for all classes to ensure a consistent average signal strength. 
Different levels of heterogeneity are introduced by setting the variance $\xi$ of the Gamma component to $0.3$, $0.6$, and $0$ for classes $1$, $2$, and $3$, respectively. 
These correspond to mild, moderate, and no non-Gaussianity in the simulated scattering behavior.

To reflect the variation in true backscattering energy among different land cover types,
we encode the mean power difference by scaling the scatter matrix.
Specifically, the scatter matrix is defined as $\bm{\Sigma} = \sigma^2 \bm{\Gamma}$,
where $\bm{\Gamma}$ is the coherence matrix with unit diagonal elements,
and $\sigma^2$ represents the class-specific mean backscatter power.

In this study, $\sigma^2$ is modeled as a reciprocal Gamma distributed random variable\footnote{We choose the reciprocal Gamma distribution for $\sigma^2$ because it is a common prior for variance or scale parameters in Bayesian models.} with probability density function
\[
f(x; \alpha, \beta) = \frac{\beta^\alpha}{\Gamma(\alpha)}\, x^{-\alpha - 1} \exp\!\big(-\frac{\beta}{x}\big),
\quad x > 0, \; \alpha, \beta > 0.
\]
We set the shape parameter $\alpha = 2$ and the rate parameter $\beta = \alpha - 1 = 1$,
which yields a distribution with unit mean but infinite variance.

To ensure realistic power contrasts among the three land cover classes,
$\sigma^2$ is set to the $10$th, $50$th, and $90$th percentiles of this reciprocal Gamma distribution,
which are approximately $0.2571$, $0.5958$, and $1.8804$, respectively.

We compare the proposed \gls{acaf} algorithm with the \gls{ccmmtf}~\cite{dong_unified_2018} method and the \gls{fashps}~\cite{jiang_fast_2015} method. \gls{ccmmtf} assesses covariance similarity under a complex Wishart model, while \gls{fashps} relies on amplitude similarity. Both \gls{ccmmtf} and \gls{fashps} are \gls{ccg}-assumption-based parametric methods. All processing in the simulated experiment is performed using a sliding window of size $11 \times 11$.

\begin{figure}[hbtp]
	\centering
	\includegraphics[width=3.5in]{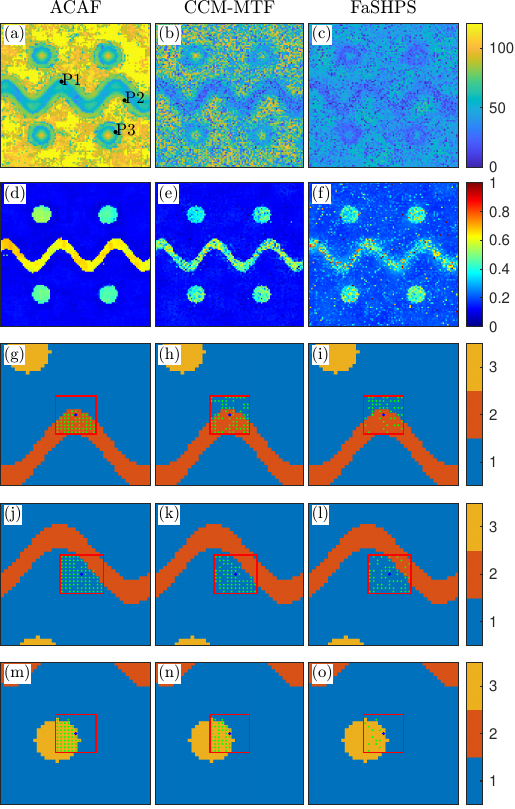} 
	\caption{
		First row: number of \glspl{shp} selected by the proposed \gls{acaf} method, the \gls{ccmmtf} method, and the \gls{fashps} method, respectively.  
		Second row: mean coherence estimated from the selected homogeneous samples using each method.  
		Remaining rows: homogeneous pixel footprints for three representative points, P1, P2, and P3 (as marked in (a)), obtained by each method, and  
		the background in these panels shows the true classification label map.
	}
	\label{fig:sim_shps} 
\end{figure}

As shown in the first row of Fig.~\ref{fig:sim_shps}, the proposed \gls{acaf} method generally selects a larger number of homogeneous pixels compared to the other two methods. The \gls{fashps} algorithm yields the fewest SHPs, while \gls{ccmmtf} lies in between.
The second row of Fig.~\ref{fig:sim_shps} shows that the coherence maps estimated by \gls{acaf} exhibit the highest spatial consistency, with only minor artifacts near the boundaries. This suggests that \gls{acaf} effectively clusters pixels with similar decorrelation behavior. In contrast, the coherence maps obtained by \gls{ccmmtf} and \gls{fashps} display more spatial noise, indicating a lower ability to distinguish between different scatterer classes.
Moreover, as expected, both \gls{ccmmtf} and \gls{fashps} perform better on the Gaussian scattering class ($\xi=0$, yellow circular region in Fig.~\ref{fig:sim_settings}(a)) than on the non-Gaussian ones with $\xi \ne 0$.

To further compare the three methods, we examine the homogeneous pixel footprints for three representative points P1, P2, and P3, as marked in Fig.~\ref{fig:sim_shps}(a). As shown in the third to fifth rows of Fig.~\ref{fig:sim_shps}, for P1, both \gls{ccmmtf} and \gls{fashps} fail to identify a set of homogeneous pixels consistent with the reference. For P2, \gls{fashps} excludes only part of the heterogeneous pixels while retaining fewer homogeneous ones. In contrast, \gls{ccmmtf} correctly excludes all heterogeneous pixels but still retains significantly fewer homogeneous pixels than \gls{acaf}. For P3, all methods successfully eliminate heterogeneous pixels, yet \gls{fashps} selects only a small portion of the homogeneous group.

Across all three cases, \gls{acaf} demonstrates the highest accuracy in distinguishing statistically homogeneous samples. These examples further illustrate that \gls{acaf} achieves a near-nominal false alarm rate while maintaining stable power, enabling it to sensitively and reliably discriminate between different interferometric coherence structures, thereby allowing more meaningful subsequent statistical inference such as phase estimation.

To evaluate the phase estimation performance of the proposed CGG-MLE and CGG-CFPL methods, we conduct comparisons using \gls{sshp} selected by the proposed \gls{acaf} approach. 
Specifically, we consider two CFPL variants, Tyler-CFPL and RegSCM-CFPL, that employ robust Tyler and regularized sample covariance estimates, respectively, as proposed in~\cite{vu_covariance_2025}. 
Additionally, we include the \gls{pta} method. As a baseline, we also evaluate a conventional DS-InSAR workflow that combines \gls{fashps} with \gls{pta} phase estimation~\cite{ferretti_new_2011}, representing the widely used conventional DS-InSAR workflow based on amplitude similarity and \gls{ccg}-based maximum likelihood estimation. 
All complex covariance matrices are regularized by shrinkage toward the identity matrix using the optimal shrinkage factor derived in~\cite{ollila_shrinking_2020}, and subsequently normalized into complex coherence matrices according to~\eqref{eq:cohm}.

\begin{figure*}[hbtp]
	\centering
	\includegraphics[width=7in]{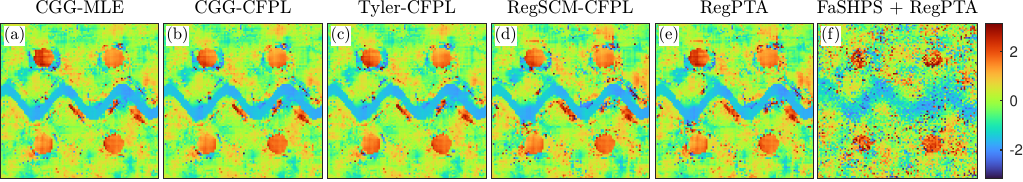} 
	\caption{Reconstructed interferograms using different phase estimation methods. 
		Panels (a)--(e) are based on homogeneous pixels selected by the proposed \gls{acaf} method, while (f) uses \gls{fashps} for homogeneous pixel selection, representing a conventional DS-InSAR workflow. 
		``RegPTA'' refers to PTA phase estimation using regularized coherence matrices.}
	\label{fig:sim_ints} 
\end{figure*}

As shown in Fig.~\ref{fig:sim_ints}, the three robust estimation-based methods (CGG-MLE, CGG-CFPL, and Tyler-CFPL) achieve the highest quality in interferometric phase reconstruction.
Meanwhile, both the matrix-matching approach (RegSCM-CFPL) and the plug-in maximum likelihood estimator (RegPTA), which rely on the CCG model, exhibit performance degradation in the simulated scenario. This highlights the advantage of robust modeling under non-Gaussian conditions. As shown in Fig.~\ref{fig:sim_ints}(f), the reconstructed interferogram from the conventional workflow displays considerable noise and blurred edges, highlighting the significant impact of homogeneous pixel selection on phase estimation quality.

\begin{figure}[hbtp]
	\centering
	\includegraphics[width=3in]{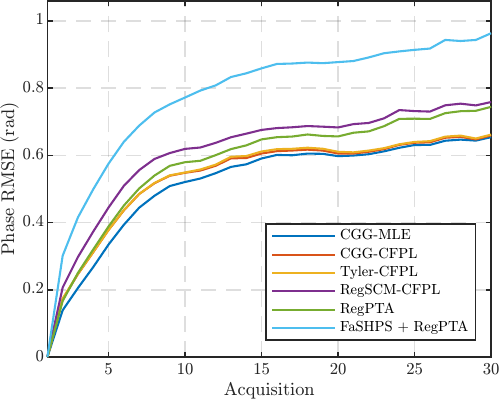} 
	\caption{
		Average phase RMSE of the reconstructed interferograms of each acquisition.
	}
	\label{fig:sim_rmse} 
\end{figure}

To further quantify the performance, we compute the average phase \gls{rmse} of the reconstructed interferograms for each acquisition. 
The results are shown in Fig.~\ref{fig:sim_rmse}. 
It can be observed that CGG-MLE achieves the highest phase estimation accuracy. 
Additionally, CGG-CFPL outperforms Tyler-CFPL by a small margin, possibly due to the CGG model partially capturing the coupling between phase and scale, whereas the Tyler estimator is entirely scale-invariant. 
Moreover, CFPL-type methods benefit from efficient iterative solutions based on the MM algorithm and avoid matrix inversion, which may enhance their robustness in high-dimensional, low-sample, or complex scenarios.

\begin{figure}[hbtp]
	\centering
	\includegraphics[width=3.5in]{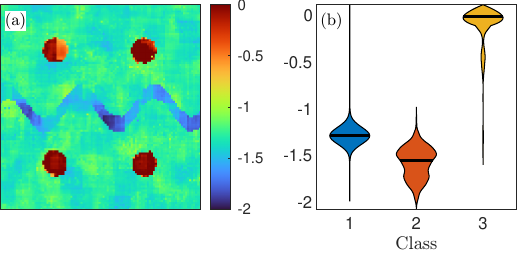} 
	\caption{
		(a) Spatial distribution of the estimated shape parameter $s$ (in $\log_{10}$ scale). 
		(b) Violin plot of $\log_{10}(s)$ values estimated under different covariance patterns. 
		Horizontal black lines indicate the median values.
	}
	\label{fig:sim_s} 
\end{figure}

Finally, Fig.~\ref{fig:sim_s} illustrates the spatial distribution of the estimated shape parameter $s$ in $\log_{10}$ scale. 
As expected, the estimated values of $s$ are lower for scatterer classes with larger heterogeneity parameter $\xi$, confirming the ability of the CGG model to reflect underlying non-Gaussian behavior. 
The violin plots in Fig.~\ref{fig:sim_s}(b) further illustrate the distribution of the estimated $\log_{10}(s)$ values across different classes. 
It can be seen that the estimates of $s$ exhibit distinct separation among the three classes, demonstrating the parameter’s practical identifiability.

\section{Tests on observed SAR data}\label{sec:real}
\subsection{Study area, data processing and results}

\begin{figure*}[hbtp]
	\centering
	\includegraphics[width=7in]{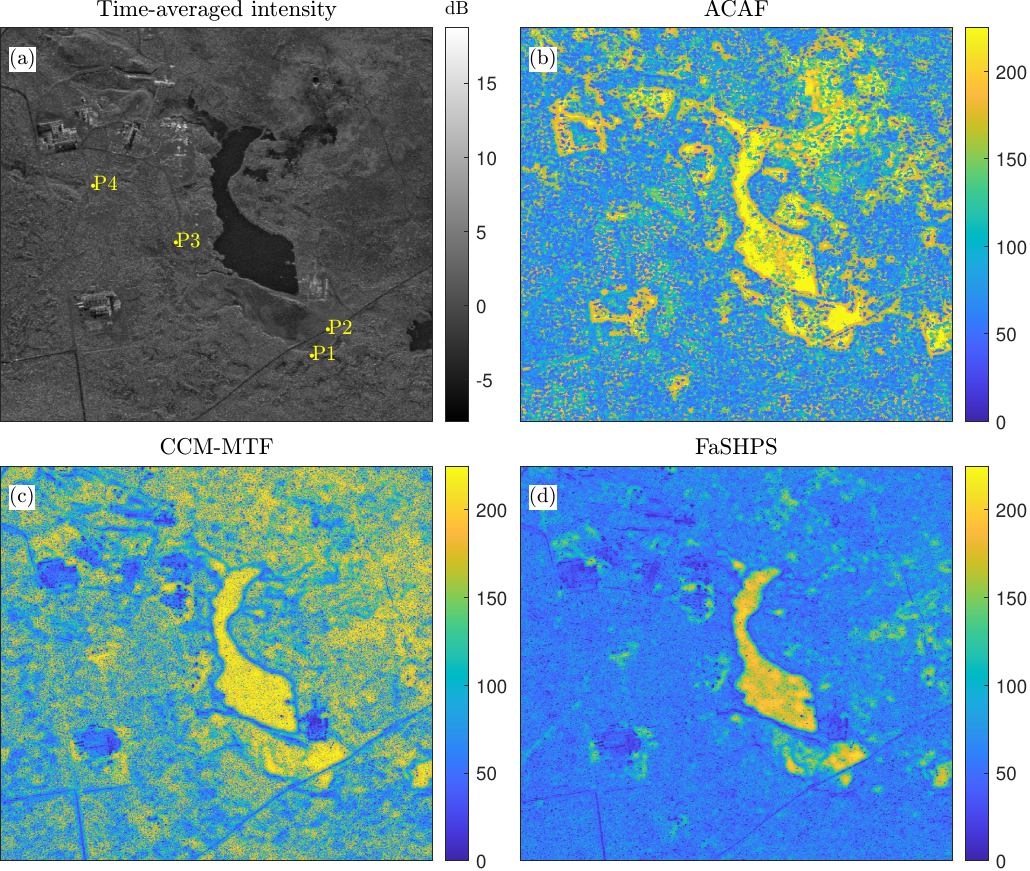} 
	\caption{
		(a) Temporal mean intensity map of the study area. 
		(b)--(d) Number of homogeneous pixels selected by \gls{acaf}, CCM-MTF and FaSHPS, respectively. 
	}		
	\label{fig:Iceland_amp_n_nshp} 
\end{figure*}

Our study area is a region in southwestern Iceland. 
Characterized by complex land cover and numerous fine-scale surface features, this area is well-suited for evaluating the performance of the proposed method. 
The dataset comprises \num{37} descending-pass TerraSAR-X StripMap images acquired in HH polarization, spanning the period from December 12, 2023, to May 1, 2025. 
Most interferometric pairs exhibit a temporal baseline of \num{11}~days, with the maximum reaching \num{33}~days. 
The temporal mean intensity image is shown in Fig.~\ref{fig:Iceland_amp_n_nshp}(a).


\begin{figure*}[hbtp]
	\centering
	\includegraphics[width=7in]{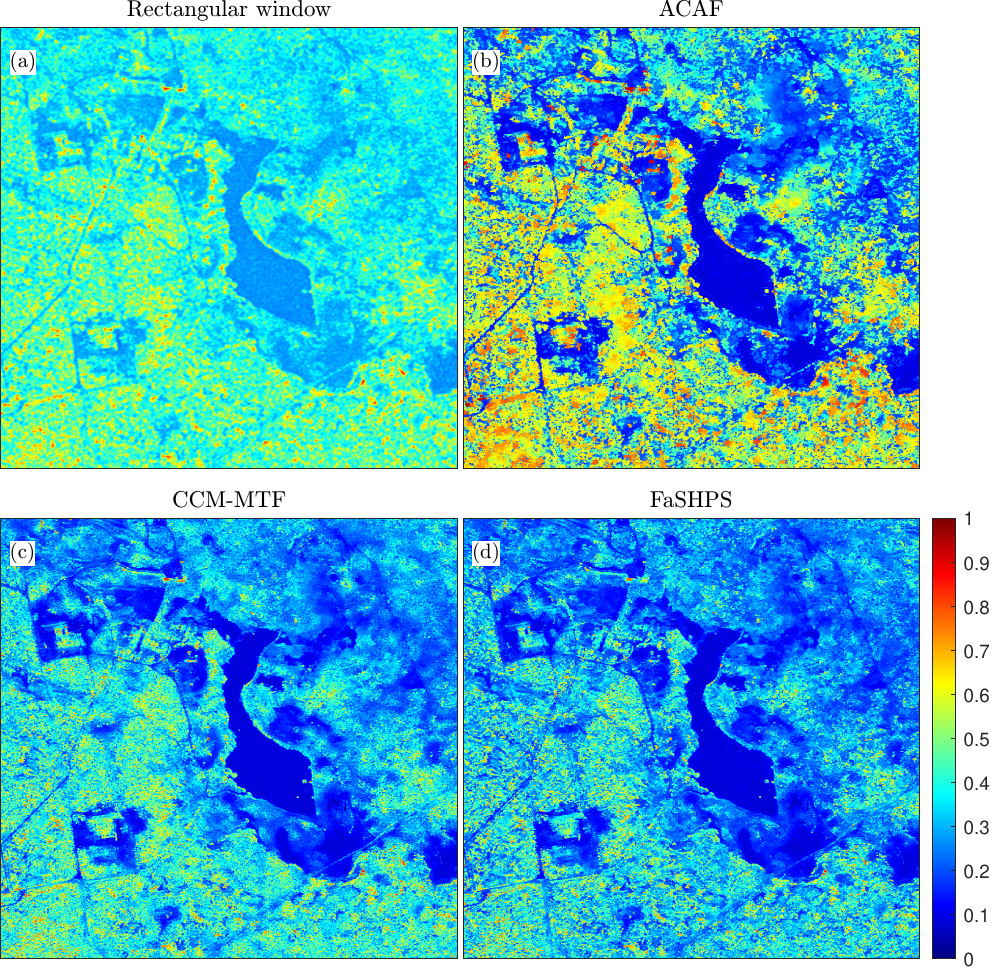} 
	\caption{(a) Mean coherence estimated using a rectangular window of size $5 \times 5$.
	(b)--(d) Mean coherence maps estimated using homogeneous pixels selectedby \gls{acaf}, CCM-MTF and FaSHPS, respectively.
	}		
	\label{fig:Iceland_meanCoh} 
\end{figure*}

We applied the proposed ACAF algorithm to identify homogeneous pixels, with CCM-MTF and FaSHPS adopted as comparative approaches.
In homogeneous pixel identification, it is common practice to retain only the connected domain (four- or eight-connected) containing the central pixel. 
To highlight differences in statistical discriminative power, the cropping procedure is applied only when the four-connected domain of the reference pixel contains more than $N+1$ samples.

The spatial distributions of the number of selected pixels are presented in Figs.~\ref{fig:Iceland_amp_n_nshp}(b)--(d). 
It can be observed that ACAF generally selects more pixels than FaSHPS, whereas the CCM-MTF method yields the largest number of selected pixels overall.

Fig.~\ref{fig:Iceland_meanCoh}(a) shows the average coherence estimated using a fixed $5 \times 5$ rectangular window, while Figs.~\ref{fig:Iceland_meanCoh}(b)--(d) show the average coherence estimated from the homogeneous pixels selected by the three methods. 
Compared to the rectangular window result, both methods exhibit sharper boundaries, demonstrating their ability to cluster statistically consistent pixels. 
The coherence maps produced by the other two comparative methods appear noisier, exhibiting more isolated discontinuities. 
In contrast, ACAF yields clearer boundaries and better spatial consistency within homogeneous regions. 
Moreover, ACAF results display stronger contrast between low- and high-coherence areas, suggesting a higher capability in distinguishing different interferometric coherence structures.

\begin{figure}[hbtp]
	\centering
	\includegraphics[width=3.5in]{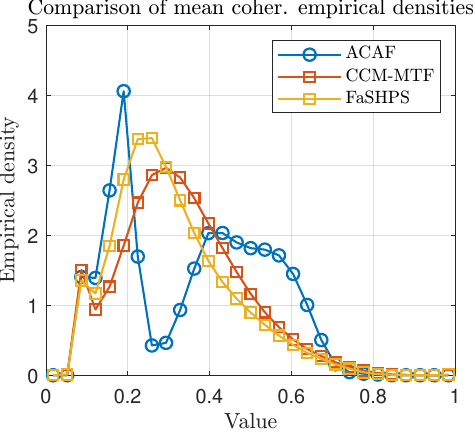} 
	\caption{
	Empirical probability density functions of the mean coherence values shown in Figs.~\ref{fig:Iceland_meanCoh}(b)--(d), corresponding to the ACAF, CCM-MTF and FaSHPS methods, respectively. 
}	
	\label{fig:Iceland_cohHist} 
\end{figure}

Fig.~\ref{fig:Iceland_cohHist} compares the empirical density curves of the mean coherence values obtained using the three methods. 
It can be observed that the result of ACAF exhibit a multimodal structure, whereas the curves of CCM-MTF and FaSHPS appear approximately unimodal. 
The multiple modes suggest that ACAF effectively separates pixels with distinct decorrelation behaviors, rather than merging them into a single homogeneous set.	

To further evaluate the performance of ACAF, four test points (P1--P4) are selected, as shown in Fig.~\ref{fig:Iceland_amp_n_nshp}(a). For each point, we present the local mean coherence maps along with the spatial footprints of the selected homogeneous pixels. First, it can be observed that the footprint of ACAF exhibits a more consistent 
spatial structure, which is more likely to correspond to objects with 
homogeneous interferometric coherence.

\begin{figure*}[hbtp]
	\centering
	\includegraphics[width=7in]{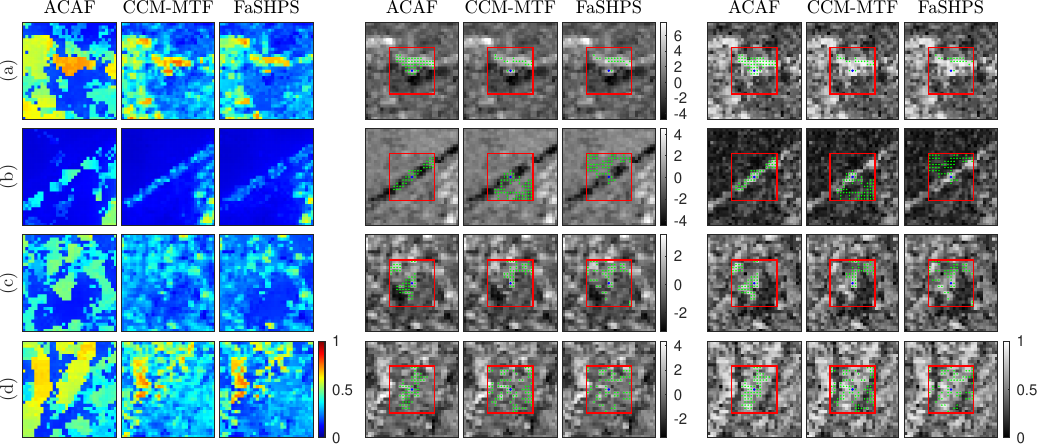} 
	\caption{
		(a)--(d) Local comparison of homogeneous pixel selection and coherence estimation for test points P1--P4 (indicated in Fig.~\ref{fig:Iceland_amp_n_nshp}(a)), respectively. 
		Columns 1--3 show local mean coherence maps estimated using ACAF, CCM-MTF and FaSHPS, respectively. 
		Columns 4 and 5 show homogeneous pixel footprints selected by the three methods, overlaid on the temporal mean intensity. 
		Columns 6 and 7 show the same footprints overlaid on the averaged cosine of the phase residual vectors, which could be regarded as a per-pixel phase statistic.
	}
	\label{fig:Iceland_points} 
\end{figure*}

Two types of background images are used in Fig.~\ref{fig:Iceland_points}: the temporal mean intensity image and the averaged cosine of phase residuals. The former reflects the backscattering strength of the target, while the latter serves as a single-pixel-level phase statistic, computed as follows:
\begin{equation}\label{eq:phStat}
	\frac{2}{N(N-1)} \sum_{i=1}^{N} \sum_{j>i}^{N} \cos\left( \theta_i - \theta_j - \arg(z_i{z_j}^*) \right) .
\end{equation}
Equation~\eqref{eq:phStat} can be interpreted as a single-pixel version of the posterior coherence~\cite{ferretti_new_2011}, where $z_i{z_j}^*$ replaces $\bm{\Gamma}_{ij}$. 

\begin{figure*}[hbtp]
	\centering
	\includegraphics[width=7in]{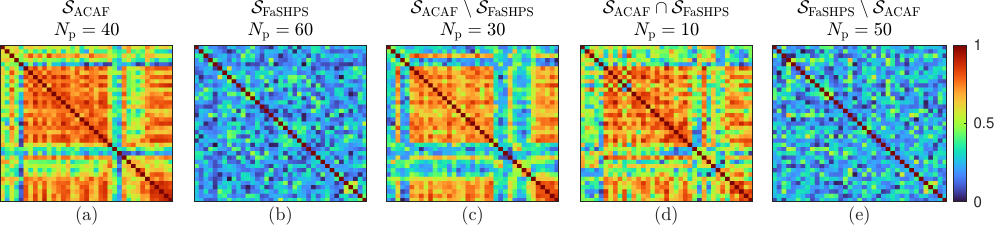} 
	\caption{Coherence magnitude matrices estimated for point P2 (see Fig.~\ref{fig:Iceland_amp_n_nshp}(a)) using different homogeneous pixel sets. From left to right: coherence estimated using (a) only the homogeneous pixels selected by ACAF, (b) only those selected by FaSHPS, (c) pixels selected exclusively by ACAF, (d) pixels selected by both ACAF and FaSHPS, and (e) pixels selected exclusively by FaSHPS. The number of pixels $N_\mathrm{p}$ used in each case is indicated above the matrix.
	}
	\label{fig:Iceland_cohms} 
\end{figure*}

For P1, the background maps in the rightmost two columns of Fig.~\ref{fig:Iceland_points}(a) indicate that both methods ensure consistent decorrelation behavior within the selected pixels. However, ACAF selects a significantly larger set of homogeneous pixels, while the pixels selected by CCM-MTF and FaSHPS appear to be a subset of those identified by ACAF. This observation may be interpreted in two ways. First, as demonstrated in the simulation experiments, the other two comparative methods may suffer from a relatively high false detection rate. Alternatively, the pixels excluded by the other two comparative methods may indeed exhibit different intensity-related statistics. ACAF, in contrast, exhibits better control over the false detection rate, possibly due to its use of the whitened statistic $t$ instead of relying solely on the temporal samples of individual pixels.

In Fig.~\ref{fig:Iceland_points}(b), the pixel sets selected by ACAF and the other two comparative methods exhibit a clear discrepancy, with only a small number of pixels falling into the intersection. The pixels selected by the other two comparative methods appear more homogeneous in terms of intensity, while ACAF includes both brighter and darker pixels. However, when using the per-pixel phase statistic as the background image, the ACAF-selected pixels demonstrate better consistency in phase-related behavior, suggesting that they may correspond to a coherent interferometric structure.

To further verify this observation, Fig.~\ref{fig:Iceland_cohms} compares the estimated coherence magnitude matrices for different pixel sets. It is evident that the matrix from the intersected pixels closely resembles the result of ACAF, indicating that distinct intensity levels can still share a common scale-invariant second-order statistical structure and could be grouped accordingly.

Compared to approaches that separate pixels based on intensity scales, the proposed ACAF method can effectively increase the homogeneous sample size. As long as the subsequent phase estimation method is either scale-invariant or can flexibly adapt to different intensity distributions, the resulting phase estimates could remain consistent.

Figs.~\ref{fig:Iceland_points}(c) and (d) show that the pixel sets selected by all three methods 
do not exhibit noticeable intensity inconsistencies. 
However, when viewed against the phase statistic background, the ACAF-selected pixels appear more consistent. 
This highlights a common limitation of intensity-based SHP selection methods, such as FaSHPS, in non-urban areas: their limited ability to distinguish scatterers that exhibit similar intensity magnitudes 
but different decorrelation behaviors~\cite{yao_phase-based_2024}.

However, experimental observations reveal that even the CCM-MTF method, which jointly exploits both intensity and phase information, may fail to fully discriminate between different land-cover types under the above scenarios. 
Possible reasons include the following. 
First, the CCM-MTF method estimates the complex covariance matrix in the temporal domain using mutually non-overlapping image pairs with the shortest time baselines. 
While this strategy effectively ensures the operability of the generalized likelihood ratio test, preserves spatial resolution, and avoids the potential instabilities of local window-based estimation, it remains essentially an approximation. 
When the primary differences in the coherence matrix are reflected in off-diagonal elements far from the main diagonal, 
or when the first-order lag autocorrelation of the observed signal exhibits pronounced second-order nonstationarity, this approximation may reduce the method’s discriminative power. 
This highlights the core challenge in statistical homogeneous pixel identification for InSAR, namely, the inherent dependence of statistical calculations on already known homogeneous samples. 
Second, although the observed signals may share similar intensity levels, real-world data do not always strictly follow the complex circular Gaussian (CCG) model. 
As illustrated by the simulation results, such statistical deviations can to some extent impair the effectiveness of parametric approaches built upon the CCG assumption.

In contrast, the proposed ACAF method focuses on modeling the consistency of the angular structure of the signal. By combining the whitened quadratic-form statistic with an inversion strategy, it circumvents the approximations and instabilities associated with temporal-domain sampling or local window-based estimation, and is not significantly affected by non-Rayleigh amplitude fluctuations. Consequently, ACAF demonstrates superior robustness and applicability.

\begin{figure}[hbtp]
	\centering
	\includegraphics[width=3.5in]{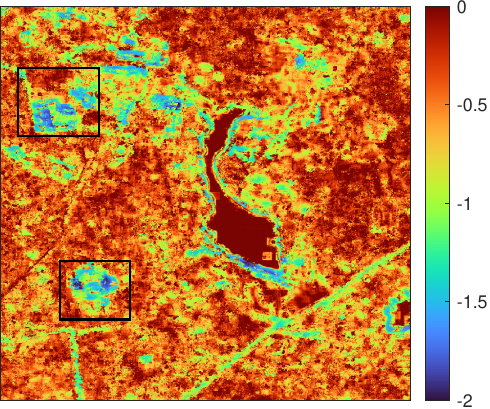} 
	\caption{
		Spatial distribution of the estimated shape parameter $s$ (in $\log_{10}$ scale). 
		Regions enclosed by black boxes exhibit noticeable non-Gaussian behavior, as indicated by lower estimated values of $s$.
	}
	\label{fig:Iceland_s} 
\end{figure}

\begin{figure*}[hbtp]
	\centering
	\includegraphics[width=7in]{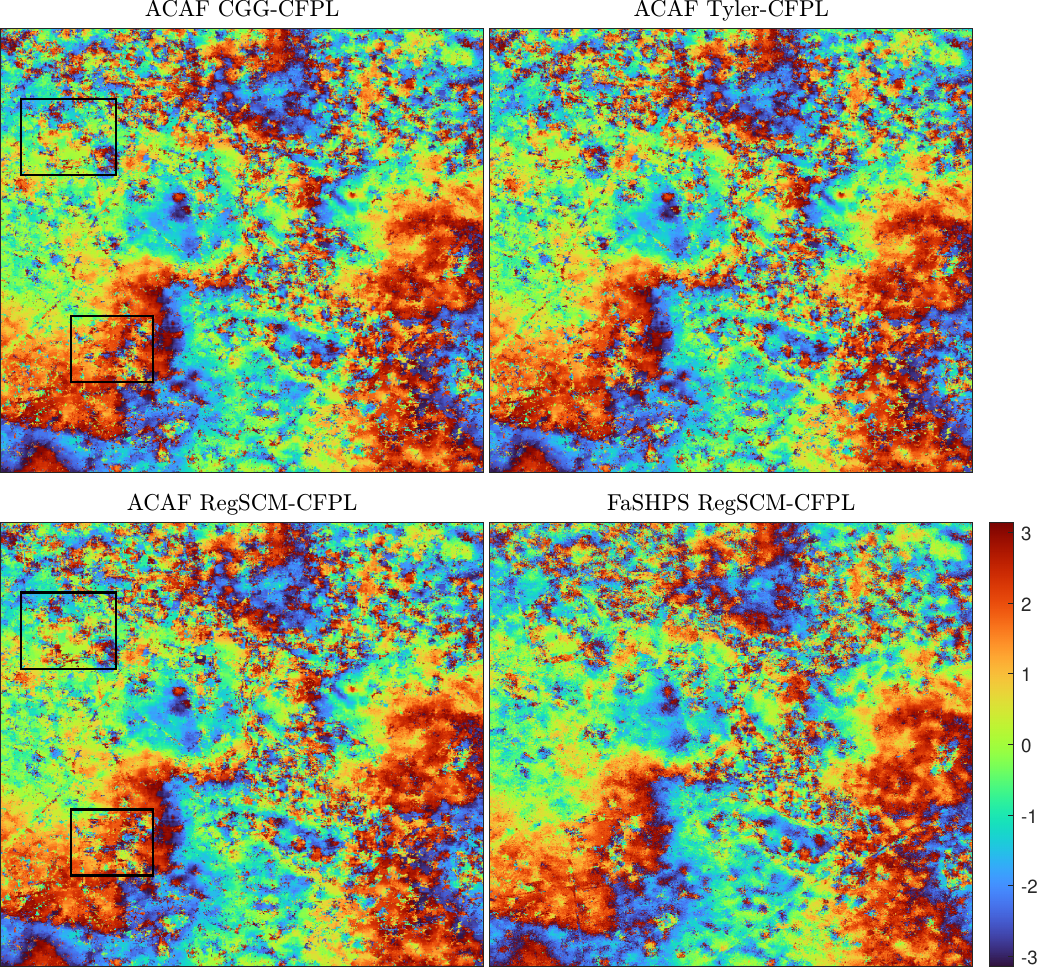} 
	\caption{
		Interferograms of the longest temporal baseline reconstructed using different phase estimation workflows: 
		(top left) ACAF + CGG-CFPL, (top right) ACAF + Tyler-CFPL, (bottom left) ACAF + RegSCM-CFPL, and (bottom right) FaSHPS + RegSCM-CFPL.
	}
	\label{fig:Iceland_ints} 
\end{figure*}

\begin{figure*}[hbtp]
	\centering
	\includegraphics[width=7in]{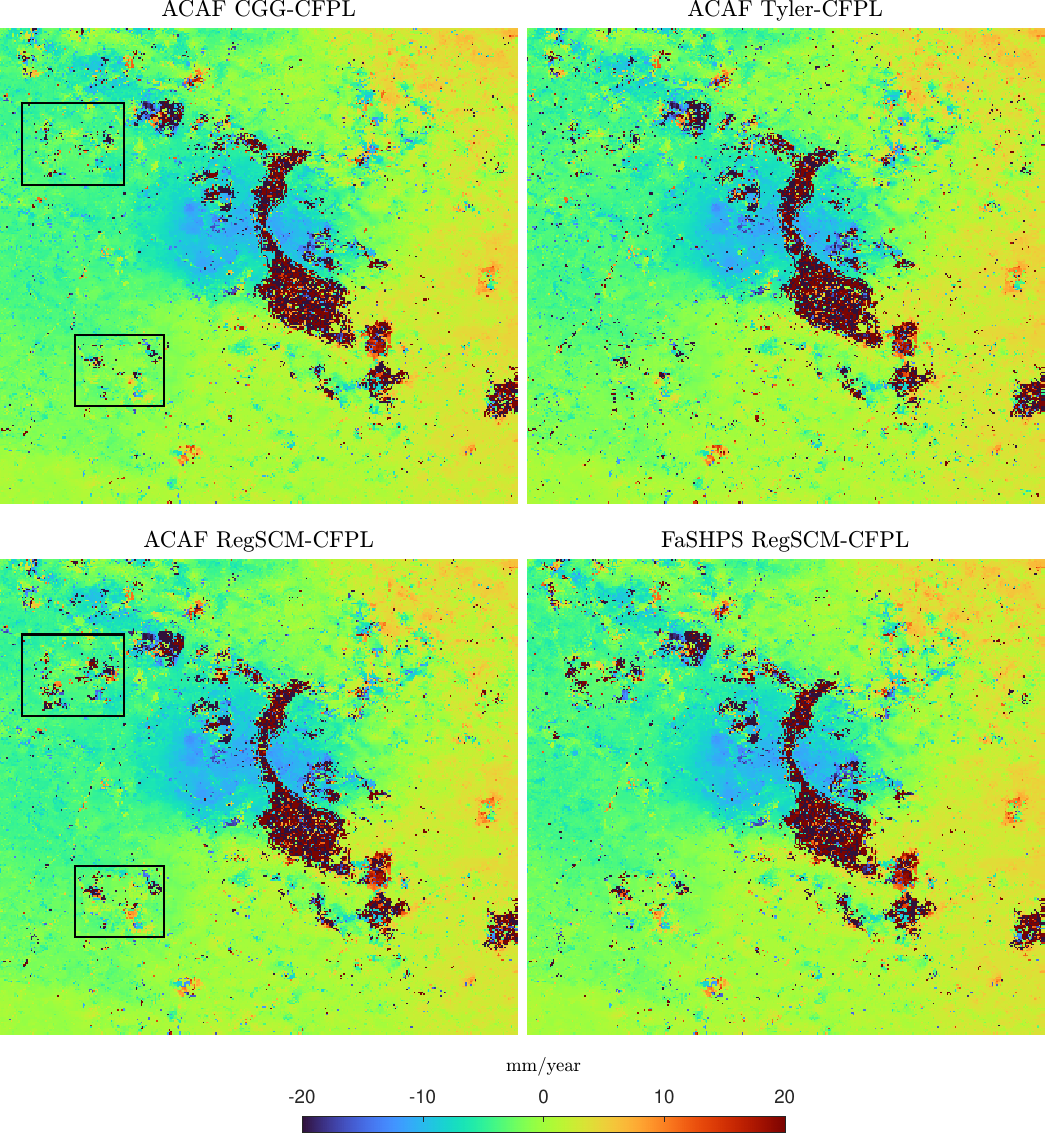} 	
	\caption{
		Estimated linear deformation rates (in mm/year) using different phase estimation workflows: 
		(top left) ACAF + CGG-CFPL, (top right) ACAF + Tyler-CFPL, (bottom left) ACAF + RegSCM-CFPL, and (bottom right) FaSHPS + RegSCM-CFPL. 
		The black boxes correspond to the regions highlighted in Fig.~\ref{fig:Iceland_s}, which exhibit notable non-Gaussian characteristics.
	}
	\label{fig:Iceland_defo} 
\end{figure*}

We subsequently conducted phase estimation experiments. Fig.\ref{fig:Iceland_s} illustrates the spatial distribution of the CGG shape parameter $s$. The black-framed area shows a clear deviation from Gaussianity and corresponds to the urban region highlighted in the yellow box in Fig.\ref{fig:Iceland_amp_n_nshp}(a), whereas natural surfaces generally exhibit less pronounced non-Gaussian behavior.

Fig.~\ref{fig:Iceland_ints} compares the reconstructed interferograms with the longest temporal baseline obtained from four different phase estimation workflows: ACAF + CGG-CFPL, ACAF + Tyler-CFPL, ACAF + RegSCM-CFPL, and FaSHPS + RegSCM-CFPL. All phase estimation methods operate under a unified and robust framework of complex coherence matrix matching~\cite{vu_covariance_2025}, and the optimization is performed using the efficient MM algorithm. This unified setting ensures that the differences among methods arise solely from the probabilistic models used for coherence matrix estimation—namely, \gls{cgg}, \gls{cacg}, and \gls{ccg}—and are not influenced by other factors such as ill-conditioning, as matrix inversion is avoided.

All coherence matrices are regularized by shrinkage toward the identity matrix to reduce estimation MSE, with the optimal shrinkage coefficient computed based on the assumed distribution model~\cite{ollila_shrinking_2020}. It is worth noting that under this configuration, RegSCM-CFPL can be interpreted as a variant of LaMIE~\cite{bai_lamie_2023}, differing only in its regularization strategy and the optimization algorithm employed.

As observed, the ACAF-based results exhibit smoother phase patterns in high-coherence regions and preserve noise characteristics in low-coherence areas, indicating effective homogeneous pixel clustering. Among the three estimation methods applied to ACAF-selected samples, the reconstructed interferograms are highly consistent across most regions. In contrast, within the non-Gaussian black-box region, the SCM-based method produces noticeable deviations from the other two robust-model-based approaches.

To further clarify the phase estimation performance of different methods, we inverted the linear deformation rates over the study area. All deformation inversions were performed using the same set of parameters for fairness. As shown in Fig.~\ref{fig:Iceland_defo}, the two robust methods based on the CGG and Tyler estimators exhibit superior performance in the non-Gaussian region.

However, in other areas, the method based on the Tyler covariance estimator introduces more noise. This may suggest that although the Tyler estimator offers robustness, its complete disregard for scale information leads to reduced efficiency. In contrast, the CGG model achieves a better trade-off between efficiency and robustness by adapting to varying intensity distributions through its shape parameter~$s$.

\subsection{Discussion}

Scale-invariant second-order statistics are central to interferometric phase estimation because they represent the angular structure and directly govern the statistical weighting of interferometric phases~\cite{cao_mathematical_2015}. In real-world scenarios, different surface targets can exhibit variations in both signal magnitude and angular structure due to their intrinsic electromagnetic scattering properties. Conventional homogeneous pixel selection methods based on amplitude similarity focus solely on signal magnitude, which tends to ensure consistency in intensity but may overlook underlying decorrelation mechanism diversity.

In contrast, the proposed ACAF method captures angular structural similarity independently of the overall energy scale. 
This makes it theoretically more suitable for supporting consistent phase-related statistical inference. 
Our real-world data experiments reveal that ACAF successfully handles cases that pose challenges for amplitude-based methods: 
(1)~pixels with differing angular structures but coincidentally similar intensity levels (e.g., point P3 in Fig.~\ref{fig:Iceland_amp_n_nshp}(a)), and 
(2)~pixels with irregular intensity patterns but consistent angular structure (e.g., point P2 in Fig.~\ref{fig:Iceland_amp_n_nshp}(a)). 
These real-world observations highlight the practical value and significance of the proposed Shape-to-Scale framework.

Despite ACAF not relying on scale information, we observe that strong non-Gaussianity is not widespread across the scene. This aligns with physical expectations. In practice, many natural terrains exhibit approximately Gaussian behavior rather than pronounced heavy-tailed scattering. This property does not hinder the performance of ACAF. On the contrary, the observed alignment with Gaussian-like behavior further confirms the physical consistency of the homogeneous pixels selected by ACAF.

Finally, ACAF can also serve as a complementary approach to amplitude-based methods by enhancing detection power in scenarios where intensity differences are subtle but angular variation is present. Such combinations can be valuable for tasks that require simultaneous and precise discrimination of both scale and correlation structure, such as fine-grained classification or change detection.

\section{Conclusion}\label{sec:conclusion}
In this paper, we proposed Shape-to-Scale, a novel framework for physically consistent DS-InSAR processing. This framework is centered around scale-invariant second-order statistics, which characterize the angular scattering structure independent of signal magnitude. To our knowledge, this is the first work to introduce a parametric method for SHP selection based on angular structural consistency (ACAF). In addition, a CGG-based \gls{pl} approach is introduced to balance robustness and efficiency in the subsequent phase estimation stage.

Extensive experiments on both simulated and real SAR datasets demonstrate that the proposed approach achieves superior performance in terms of coherence structure clustering accuracy and phase estimation quality.
The Shape-to-Scale framework offers a unified and physically interpretable solution for DS-InSAR processing and shows consistently strong performance in real-world applications.
This work provides a new perspective on robust and consistent statistical inference in heterogeneous and complex scattering environments, and offers useful insights for advancing high-resolution SAR time series analysis.

\section{Acknowledgment}
The TerraSAR-X data were provided by the German Aerospace Center (DLR) via the super sites initiative.

\small
\bibliographystyle{IEEEtranN}
\bibliography{refs}

\appendix
\section{Analysis of the CACG-Based Test Statistic}\label{appendix:t_statistic_analysis}
This appendix provides a justification for the reliability of using a right-tail test to reject low-quality heterogeneous pixels.  
In the proposed \gls{acaf}, a common normalized scatter matrix, denoted by ${\bm \Sigma}_{\rm ref}$, is employed to compute the test statistic 
\[
t = \widetilde{\mathbf{z}}^\dagger {\bm \Sigma}_{\rm ref}^{-1} \widetilde{\mathbf{z}}
\]
for each pixel in the local window.  
We assume that the normalized temporal vector $\widetilde{\mathbf{z}}$ of a given pixel admits the representation 
\[
\widetilde{\mathbf{z}} = \frac{\mathbf{A} \mathbf{u}}{\|\mathbf{A} \mathbf{u}\|}, 
\quad \text{with} \quad 
\mathbf{A} \mathbf{A}^\dagger = \bm \Sigma,
\]
where $\mathbf{u} \in \mathbbm{C}^N$ is a unit-norm random vector uniformly distributed on the complex unit sphere.  
By construction, the normalized covariance satisfies 
\[
\mathbbm{E}[\widetilde{\mathbf{z}}\widetilde{\mathbf{z}}^\dagger] 
= \frac{\bm \Sigma}{\operatorname{tr}(\bm \Sigma)}.
\]
Note that the true scatter matrix $\bm \Sigma$ for each candidate pixel may or may not coincide with the reference matrix ${\bm \Sigma}_{\rm ref}$ used for whitening.

We can obtain that
\begin{equation}
	t 
	= \widetilde{\mathbf z}^\dagger \bm{\Sigma}_{\mathrm{ref}}^{-1} \widetilde{\mathbf z} 
	= \frac{(\mathbf A \mathbf u)^\dagger \bm{\Sigma}_{\mathrm{ref}}^{-1} (\mathbf A \mathbf u)}
	{(\mathbf A \mathbf u)^\dagger (\mathbf A \mathbf u)} 
	= \frac{\mathbf u^\dagger \mathbf A^\dagger \bm{\Sigma}_{\mathrm{ref}}^{-1} \mathbf A \mathbf u}
	{\mathbf u^\dagger \bm{\Sigma} \mathbf u}.
\end{equation}
\begin{align}
	\mathbbm{E}[t]&=\mathbbm{E}[\operatorname{tr}(\widetilde{\mathbf z}^\dagger \bm{\Sigma}_{\mathrm{ref}}^{-1} \widetilde{\mathbf z} )]\notag \\
	&=\mathbbm{E}[\operatorname{tr}(\bm{\Sigma}_{\mathrm{ref}}^{-1} \widetilde{\mathbf z} \widetilde{\mathbf z}^\dagger )]=\frac{ \operatorname{tr}\left(\bm{\Sigma}_{\mathrm{ref}}^{-1}{\bm\Sigma}\right) }
	{\operatorname{tr}\left({\bm\Sigma}\right)}.
\end{align}
When $\bm \Sigma = {\bm \Sigma}_{\rm ref}$, $t$ and its expectation $\mathbbm{E}[t]$ simplify to
\begin{align}
	t &=\frac{1}
	{\mathbf u^\dagger \bm{\Sigma} \mathbf u}, \text{ and }
	\mathbbm{E}[t] = \frac{N}{\operatorname{tr}(\bm{\Sigma})}, 
\end{align}
respectively.
By applying the eigen-decomposition:
\[
	\bm{\Sigma}_{\mathrm{ref}} = \mathbf{U}_{\mathrm{ref}}\bm{\Lambda}_{\mathrm{ref}}\mathbf{U}_{\mathrm{ref}}^{\dagger},\quad
	\bm{\Lambda}_{\mathrm{ref}}=\operatorname{diag}(\lambda_1,\dots,\lambda_N),\]
\[	\mathbf{U}_{\mathrm{ref}}\mathbf{U}_{\mathrm{ref}}^{\dagger}=\mathbf{I}.\]
Similarly
\begin{equation}
	\bm{\Sigma}=\mathbf{U}\bm{\Lambda}\mathbf{U}^{\dagger}, \quad \mathbf{U}\mathbf{U}^{\dagger}=\mathbf{I}.
\end{equation}
Define the rotated vector
\begin{equation}
	\mathbf{v}=\mathbf{U}^{\dagger}\mathbf{u},
\end{equation}
where $\mathbf{v}$ is uniformly distributed in $\mathbbm{C}S^{N-1}$.
Denote $\mathbf{C}=\mathbf{U^\dagger U}_{\mathrm{ref}}$.
Substituting $\mathbf v$ and $\mathbf C$ into the definition of the test statistic gives
\begin{equation}
	t=\frac{\mathbf{v}^{\dagger}\bm{\Lambda}^{1/2}\mathbf{C}\bm{\Lambda}_{\mathrm{ref}}^{-1}\mathbf{C}^{\dagger}\bm{\Lambda}^{1/2}\mathbf{v}}
	{\mathbf{v}^{\dagger}\bm{\Lambda}\mathbf{v}}.
\end{equation}

Let us analyze the behavior of the test statistic under the case of spectral flattening form given by
\begin{equation}
	\bm{\Sigma}(\varepsilon)
	= (1-\varepsilon)\,\bm{\Sigma}_{\mathrm{ref}} + \varepsilon\,\frac{\operatorname{tr}(\bm{\Sigma}_{\mathrm{ref}})}{N}\mathbf{I},
	\quad 0<\varepsilon<1 .
	\label{eq:Sigma_flatten}
\end{equation}
In this scenario, the scatter matrix reduces to the 
Equation~\eqref{eq:Sigma_flatten} gradually flattens the eigenvalue spectrum, thus modeling a DS of lower ensemble phase quality.

The test statistic simplifies to
\begin{equation}
	t(\varepsilon)=\frac{\mathbf{v}^{\dagger}\bm{\Lambda}_{\varepsilon}^{1/2}\bm{\Lambda}_{\mathrm{ref}}^{-1}\bm{\Lambda}_{\varepsilon}^{1/2}\mathbf{v}}{\mathbf{v}^{\dagger}\bm{\Lambda}_{\varepsilon}\mathbf{v}}
	= \frac{\sum_{i=1}^{N}w_i\frac{\alpha_i}{\lambda_i}}{\sum_{i=1}^{N}w_i\alpha_i},
\end{equation}
where $\alpha_i=(1-\varepsilon)\lambda_i+\varepsilon\bar{\lambda}$, $\bar{\lambda}={N}^{-1}\sum_{i=1}^N \lambda_i$, and we defined $w_i = |v_i|^2$ satisfying $\sum_i w_i=1$.

Introduce the notation
\begin{equation}
	\mu=\sum_{i=1}^{N} w_i \lambda_i, \quad m=\sum_{i=1}^{N} w_i \lambda_i^{-1}.
\end{equation}
Then the numerator and denominator of $t(\varepsilon)$ become
\begin{align}
	N(\varepsilon) &= 1 - \varepsilon + \varepsilon \bar{\lambda} m, \\
	D(\varepsilon) &= (1 - \varepsilon) \mu + \varepsilon \bar{\lambda}.
\end{align}
Differentiating $t(\varepsilon)$ with respect to $\varepsilon$ using the quotient rule, we have
\begin{equation}
	\frac{d t}{d\varepsilon}
	= \frac{N'(\varepsilon) D(\varepsilon) - N(\varepsilon) D'(\varepsilon)}{D(\varepsilon)^2},
\end{equation}
where
\begin{equation}
	N'(\varepsilon) = -1 + \bar{\lambda} m, \quad D'(\varepsilon) = -\mu + \bar{\lambda}.
\end{equation}
Simplifying the numerator, we obtain
\begin{equation}
	N'(\varepsilon) D(\varepsilon) - N(\varepsilon) D'(\varepsilon) = \bar{\lambda}(m \mu - 1).
\end{equation}
Applying the harmonic-arithmetic mean inequality, we know
\begin{equation}	
	m \mu \geq 1,
\end{equation}
with equality if and only if all eigenvalues are identical. Physically, identical eigenvalues correspond to a completely decorrelated scenario, and the corresponding coherence matrix $\bm{\Gamma}=\mathbf{I}$.
Hence,
\begin{equation}
	\frac{d t}{d\varepsilon} \geq 0,
\end{equation}
establishing the monotonic increase of $t(\varepsilon)$ with respect to $\varepsilon$. Accordingly, if ${\bm\Sigma}_{\mathrm{ref}}$ represents a highly coherent
ensemble (i.e., has a sharply peaked eigenvalue spectrum), any degradation of
ensemble phase quality in a candidate pixel---modelled by larger
$\varepsilon$---will force its statistic into the right tail, allowing such
low-quality heterogeneous pixels to be consistently rejected by the
right-sided test.
\begin{IEEEbiography}
	[{\includegraphics[width=.83in,clip]{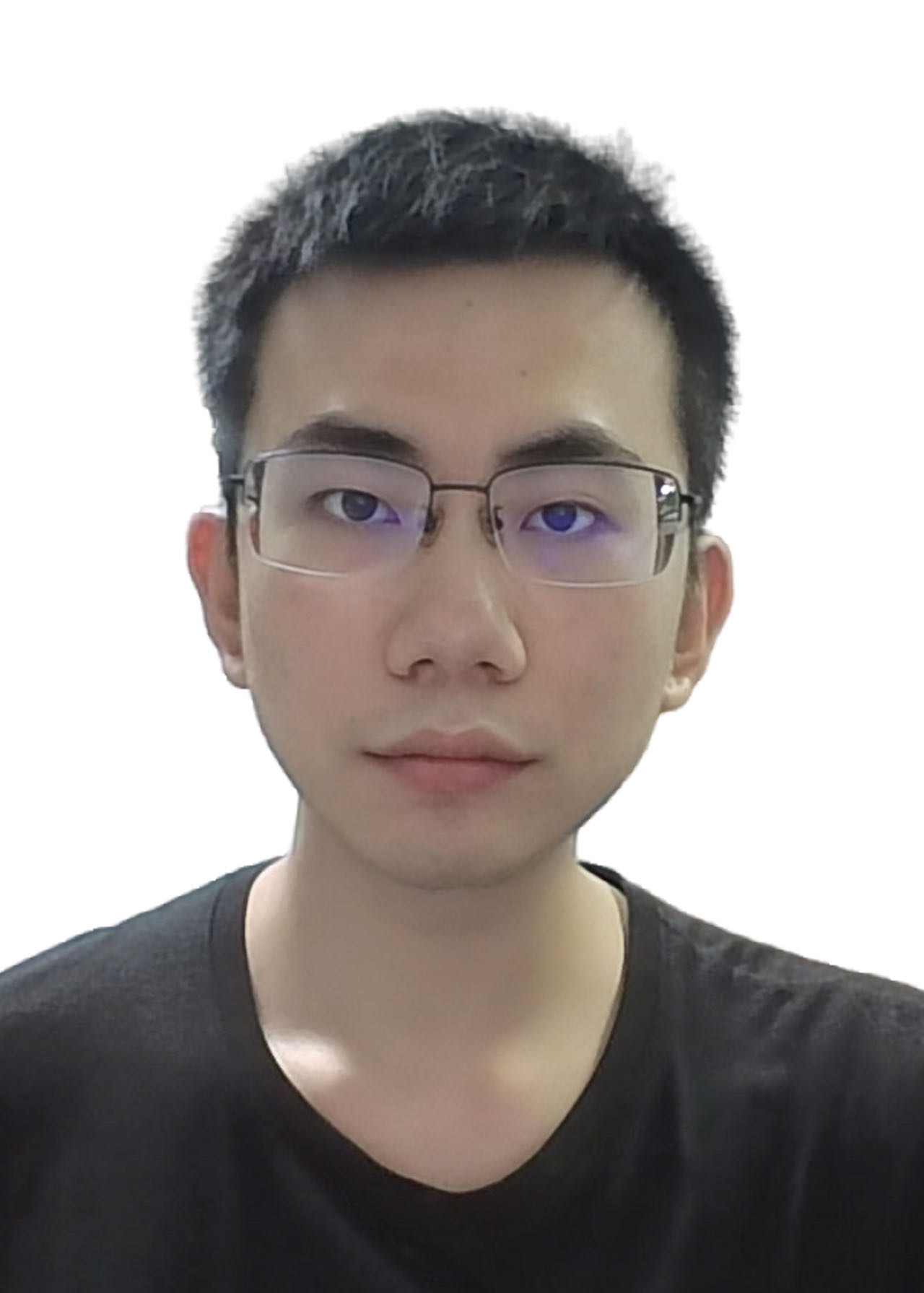}}]{Shuyi Yao} 
	received his B.S. degree in surveying and mapping engineering and his M.S. degree in photogrammetry and remote sensing from China University of Mining and Technology, Beijing, China, in 2018 and 2021, respectively. He is pursuing his Ph.D.  degree in photogrammetry and remote sensing with the State Key Laboratory of Information Engineering in Surveying, Mapping, and Remote Sensing, Wuhan University, Wuhan 430079, China. He is also a joint Ph.D. student at Victoria University of Wellington, Wellington, New Zealand. His research interests include statistical modeling and robust estimation of interferometric synthetic aperture radar signals over distributed scatterers as well as geophysical inversion. He is a Student Member of IEEE.
\end{IEEEbiography}
\begin{IEEEbiography}
	[{\includegraphics[width=.83in,clip]{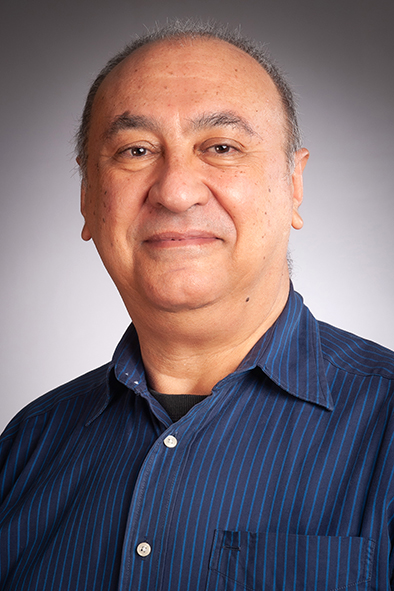}}]{Alejandro C. Frery} 
	(S'91--M'94--SM'14--F'25) was born in Mendoza, Argentina, in 1960. In 1983, he received a B.Sc. in Electronic and Electrical Engineering from the Universidad de Mendoza. His M.Sc. degree was in Applied Mathematics (Statistics) from the Instituto de Matemática Pura e Aplicada (IMPA, Rio de Janeiro, 1990), and his Ph.D. degree was in Applied Computing from the Instituto Nacional de Pesquisas Espaciais (INPE, S\~{a}o José dos Campos, Brazil, 1993). His research interests are information visualisation, statistical computing, and stochastic modelling.
	Since May 2020, he has been a Statistics and Data Science Professor at the School of Mathematics and Statistics, Victoria University of Wellington, New Zealand.
	Prof.\ Frery is also a member of the International Society for Photogrammetry and Remote Sensing (ISPRS), the New Zealand Statistical Association, and the Statistical Society of Australia. 
	In 2018, he received the IEEE GRSS Regional Leader Award. 
	After serving as Associate Editor for over five years, Prof.\ Frery was the Editor-in-Chief of the IEEE Geoscience and Remote Sensing Letters from 2014 to 2018. 
	He was an IEEE Geoscience and Remote Sensing Society (GRSS) Distinguished Lecturer from 2015 to 2019. 
	He served as this Society's AdCom (Advisory Committee) member, in charge of Future Publications, Plagiarism and Regional Symposia from 2019 to 2022. 
	Since 2023, he has been IEEE GRSS Vice-President of Publications. 
\end{IEEEbiography}
\begin{IEEEbiography}
	[{\includegraphics[width=.83in,clip]{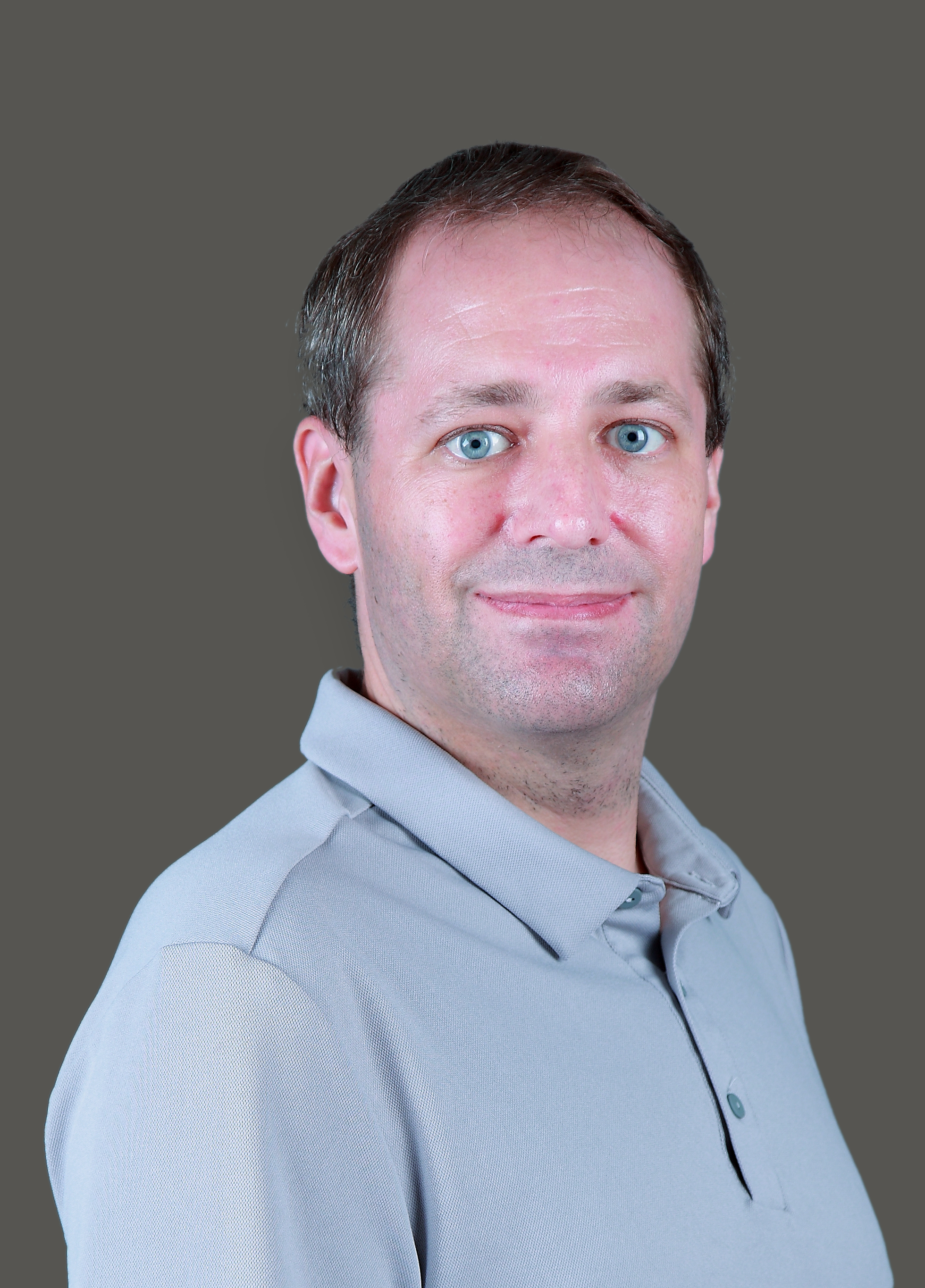}}]{Timo Balz} 
	received his diploma degree in geography (Dipl.-Geogr.) and his doctoral degree (Dr.-Ing.) in aerospace engineering and geodesy from the University of Stuttgart, Stuttgart, Germany, in 2001 and 2007, respectively. From fall 2001 to the end of 2007, he was a research assistant with the Institute for Photogrammetry, University of Stuttgart. From 2008 to 2010, he was a postdoctoral research fellow with the State Key Laboratory of Information Engineering in Surveying, Mapping, and Remote Sensing (LIESMARS), Wuhan University. From 2010 to 2015, he was an associate professor for radar remote sensing with LIESMARS. Since 2015, he has been a full professor with LIESMARS, and since 2021, he has been the vice director of the International Academy of Geoinformation, Wuhan University, Wuhan 430079, China. His research interests include surface motion estimation with synthetic aperture radar (SAR), data visualization, SAR geodesy, and the use of SAR data to support archaeological prospections. He is a Senior Member of IEEE.
\end{IEEEbiography}
\end{document}